\documentclass[mathpazo]{csam}

\usepackage{bm}
\usepackage{xcolor}
\usepackage{cite}
\usepackage[noend]{algpseudocode}
\usepackage[pagewise]{lineno}
\usepackage{algorithmicx,algorithm,setspace,subfigure}
\usepackage{booktabs,multirow}
  
\usepackage{appendix}
\DeclareMathOperator*{\mmin}{minimize}
 
\newcommand{\mbbR}{\mathbb{R}} 
 
\newcommand{\mbbE}{\mathbb{E}}

\newcommand{\bdw}{\boldsymbol{w}} 
\newcommand{\bdr}{\boldsymbol{r}}

\newcommand{\bda}{\boldsymbol{\alpha}}

\newcommand{\mcX}{\mathcal{X}}

\newcommand{\mcY}{\mathcal{Y}}

\newenvironment{proofs}{{\indent \it \textbf{Proof}:\quad}}{\hfill $\blacksquare$\par}
\makeatother
\allowdisplaybreaks[4]

\begin{document}
\title{A Constrained BA Algorithm for Rate-Distortion \\ and Distortion-Rate Functions}


\author[L. Chen et~al.]{Lingyi Chen\affil{1}, Shitong Wu\affil{1}, Wenhao Ye\affil{1} \\ Huihui Wu\affil{2}\comma\corrauth, Wenyi Zhang\affil{4}, Hao Wu\affil{1} ~and Bo Bai\affil{3}}
\address{
        \affilnum{1}\ Department of Mathematical Sciences, Tsinghua University, Beijing 100084, P.R. China. \\
        \affilnum{2}\ Yangtze Delta Region Institute (Huzhou), University of Electronic Science and Technology of China, Huzhou, Zhejiang, 313000, P.R. China. \\
        \affilnum{3}\ Theory Lab, Center Research Institute, 2012 Labs, Huawei Technologies Co. Ltd., Hong Kong SAR. \\ 
        \affilnum{4}\ Department of Electronic Engineering and Information Science, University of Science and Technology of China, Hefei, Anhui 230027, P.R. China.
        }         
\emails{{\tt huihui.wu@ieee.org} (H. Wu).}

\begin{abstract}
The Blahut-Arimoto (BA) algorithm has played a fundamental role in the numerical computation of rate-distortion (RD) functions. 
This algorithm possesses a desirable monotonic convergence property by alternatively minimizing its Lagrangian with a fixed multiplier. 
In this paper, we propose a novel modification of the BA algorithm, wherein the multiplier is updated through a one-dimensional root-finding step using a monotonic univariate function, efficiently implemented by Newton's method in each iteration. 
Consequently, the modified algorithm directly computes the RD function for a given target distortion, without exploring the entire RD curve as in the original BA algorithm. 
Moreover, this modification presents a versatile framework, applicable to a wide range of problems, including the computation of distortion-rate (DR) functions. 
Theoretical analysis shows that the outputs of the modified algorithms still converge to the solutions of the RD and DR functions with rate $O(1/n)$, where $n$ is the number of iterations. 
Additionally, these algorithms provide $\varepsilon$-approximation solutions with $O\left(\frac{MN\log N}{\varepsilon}(1+\log |\log \varepsilon|)\right)$ arithmetic operations, where $M,N$ are the sizes of source and reproduced alphabets respectively. 
Numerical experiments demonstrate that the modified algorithms exhibit significant acceleration compared with the original BA algorithms and showcase commendable performance across classical source distributions such as discretized Gaussian, Laplacian and uniform sources. 
\end{abstract}


\maketitle


\section{Introduction}
The rate-distortion (RD) theory, first introduced by Shannon \cite{shannon1948mathematical, shannon1959coding}, characterizes the fundamental trade-off between the rate and the distortion in lossy compression \cite{berger1971, berger98}. 
Nowadays, the application scope of RD theory is extensive and far-reaching. 
Notably, it underpins critical technologies and standards, including image and video compression standards such as JPEG, MPEG, and H.264 standards \cite{model_skodras2001jpeg,taubman2002jpeg2000,le1991mpeg,model_wiegand2003overview}. 
Additionally, the RD theory and its relevance also extend to machine learning, particularly in the realm of quantization design \cite{balle2017end,jpegai_whitepaper} in learned lossy compression models. 
This multifaceted utility highlights the importance of RD theory in modern technology and research. 

The computation of the RD function is of great interest and stands as a central concern within the field of RD theory.
The RD function is obtained by minimizing the mutual information between the source and the reproduction subject to an average distortion constraint. 
Formally stated, given a source random variable $X\in\mcX$ with probability distribution $P_{X}$, a reproduction $Y\in\mcY$ and a distortion measure $d: \mcX\times\mcY\rightarrow \mathbb{R}^+=[0, \infty)$, the RD function is defined as\cite{book_element,berger1971}, 
\begin{equation}\label{rate_dis_tor_def}
R(D):=\min_{P_{Y|X}} I(X ; Y) \quad \text { s.t. } \quad \mbbE_{P_{XY}}\left[d(X,Y)\right]\leq D.
\end{equation}
Here, $Y\in\mcY$ is understood as the compressed representation, i.e., the lossy reproduction of $X$. The mutual information $I(X; Y)$ is minimized with respect to the conditional probability distribution $P_{Y|X}(y|x)$, and the expected distortion is constrained by a target distortion threshold $D$.
To date, the prevailing numerical method for computing the RD function has been the Blahut-Arimoto (BA) algorithm \cite{blahut1972computation, arimoto1972algorithm}, which minimizes the following RD Lagrangian,
\begin{equation} \label{RD_Larg}
\mathcal{L}_{\mathrm{RD}}^{\lambda}(P_{Y|X}):=I(X ; Y)+\lambda~\mbbE_{P_{XY}}\left[d(X,Y)\right],
\end{equation}
for each fixed multiplier $\lambda\in\mbbR^{+}$. Geometrically, $\lambda$ corresponds to the slope of the tangent line of the RD curve. 
The tangent point $(D_{\lambda}, R_{\lambda})$ associated with each fixed $\lambda\in\mbbR^{+}$ is computed by the optimal solution $P^{*}_{Y|X}$ in the RD Lagrangian \eqref{RD_Larg}, i.e., $R_{\lambda}=I(X ; Y^{*})$ and $D_{\lambda}=\mbbE\left[d(X,Y^{*})\right]$ where $Y^{*}$ denotes the random variable generated by $P_X P^{*}_{Y|X}$.
Hence, keeping varying the slope, i.e., the multiplier $\lambda$, we can ``sweep out'' the entire RD curve.

The BA algorithm is an iterative procedure built upon this geometric view, which has been shown to converge to the RD function (e.g., \cite{BA_csiszar1974computation, csiszar11}).
Various extensions (e.g., \cite{dupuis2004}) and acceleration techniques (e.g., \cite{sayir2000, matz2004, yu2010} for channel capacity) for the BA algorithm have been studied. 
In addition, a mapping approach \cite{rose94} has been proposed mainly for computing the RD curve for sources with continuous amplitudes.
However, a common theme throughout these studies has been fixing the multiplier $\lambda$ throughout iterations when minimizing the RD Lagrangian.
As a consequence, to obtain the RD function $R(D)$ for a given target distortion $D$, these algorithms have to explore the RD curve to search for the corresponding slope $\lambda$. Basically, a number of trials with different choices of $\lambda$ need to be conducted using an adaptive procedure, and eventually the desired choice of $\lambda$ corresponding to the given target distortion $D$ is obtained, as well as the corresponding $R(D)$. This process incurs a heavy computational cost and may cause numerical instability \cite{agmon2021critical,agmon2022root} when the RD curve possesses linear segments, where $\lambda$ is a constant. 
To address the aforementioned difficulty, in this paper we present a novel modification on the BA algorithm, referred to as the \textbf{C}onstrained \textbf{B}lahut-\textbf{A}rimoto (\textbf{CBA}) algorithm, revealing that the RD function can be computed directly with a given target distortion, by updating the multiplier in each iteration via additionally solving an one-dimensional monotonic function over the BA algorithm. 
This modification is partially inspired by our recent works \cite{wu2022communication, ye2022optimal}, that consider the numerical computation of bounds in information theory from an optimal transport (OT) perspective, as these problems can be formulated as regularized OT problems with additional constraints \cite{2022wth, peyre2019computational}.
A noteworthy fact is that, the additional constraints in these problems can be converted into one-dimensional root-finding problems with respect to certain monotonic univariate function, which can be efficiently solved by Newton's method with only a few iterations. 
Hence, in this paper, we update the multiplier similarly with Newton's method in each iteration, and this seemingly simple modification renders the CBA algorithm fundamentally different from the original BA algorithm, which uses a fixed multiplier throughout iterations. 
It is worth highlighting that the proposed modification is a general framework that extends beyond the scope of the current RD problem and can be applied to a wide range of scenarios, including the computation of distortion-rate (DR) functions.
Moreover, we provide a theoretical analysis of the CBA algorithm by exploring the optimal condition of each alternating direction, proving its convergence and establishing that the convergence rate is $O(1/n)$, where $n$ is the number of iterations. 
Given an arbitrary $\epsilon>0$, our proposed CBA algorithm can provide a solution with $\epsilon$-approximation to the optimal solution in objective value with arithmetic operations of $O\left(\frac{MN\log N}{\varepsilon}(1+\log |\log \varepsilon|)\right)$ for the RD and DR functions, which outperforms the existing results in the literature, including those from well-established methods \cite{hayashi2023bregman,blahut1972computation,wu2022communication}.
\begin{figure}[ht]
    \centerline{\includegraphics[width=.95\textwidth]{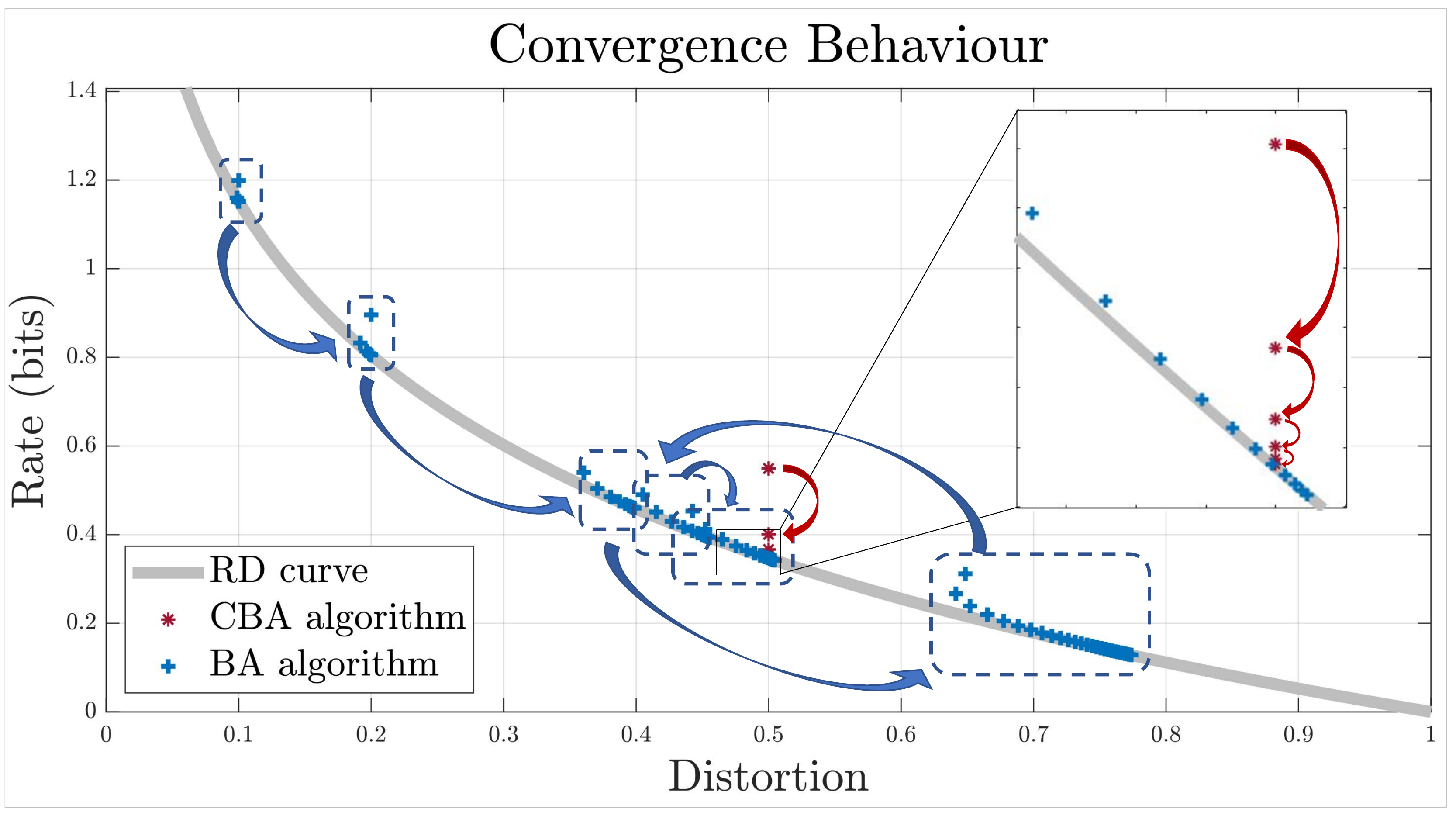}}
   \caption{Comparison between convergent trajectories of the CBA algorithm and the BA algorithm, for computing $R(D)$ with $D = 0.5$.}
   \label{Fig: converge}
\end{figure}
The difference between the CBA algorithm and the BA algorithm is graphically illustrated in Figure \ref{Fig: converge}, which displays the convergence trajectories of the two algorithms.
The updating strategy of the BA algorithm is searching for the slope $\lambda$ corresponding to the given target $D$, and for each trial on $\lambda$, its execution incurs a variable number of inner iterations.
As a result, in Figure \ref{Fig: converge}, the convergence trajectory of the BA algorithm comprises both a ``macroscopic'' external part, represented by blue arrows between blocks, and a ``microscopic'' inner part, depicted by the marks within each block.
In contrast, the CBA algorithm directly computes the RD function $R(D)$ with a given target distortion $D$, without relying on fixing the slope $\lambda$. 
Consequently, in Figure \ref{Fig: converge}, the convergence trajectory of the modified algorithm is simply a vertical array of marks.
Apparently, our numerical experiments demonstrate that when computing the RD function directly with a given target distortion, the CBA algorithm significantly accelerates the BA algorithm.
Moreover, for the case where the RD curve contains a linear segment, since there lacks a one-to-one mapping between the slopes $\lambda$ and the tangent points $(D_{\lambda}, R_{\lambda})$, the BA algorithm usually encounters numerical instability \cite{agmon2021critical} and fails to produce the entire RD curve. 
In contrast, the CBA algorithm still applies effectively, since it does not rely on the correspondence between $\lambda$ and $(D_{\lambda}, R_{\lambda})$.
It is worth noting that a recent work \cite{hayashi2023bregman}, which utilizes the Bregman divergence based EM (Expectation-Maximum) algorithm for computing classical and quantum rate distortion functions, has also obtained a similar algorithm in the classical rate distortion case.
In contrast to the aforementioned work based on the EM framework \cite{fujimoto2007em}, our approach revolves around analyzing the Lagrangian, leading to an entirely different framework for deriving the algorithm and analyzing convergence properties. 
Furthermore, our approach has a broader applicability, easily extending to various problems, including the computation of DR functions, which poses difficulties for EM-type algorithms primarily because the objective function cannot be expressed as alternating projections into two convex sets.
Through our Lagrangian analysis, we can easily derive a single-variable monotonous equation as in \cite{wu2022communication}, enabling the application of the Newton's method for root-finding problems, yielding superior numerical results in Section \ref{Sec_4_numerical}.
The numerical outcomes highlight the notable acceleration achieved by our proposed CBA algorithm, with speed-up ratios reaching up to $10$ when compared to existing algorithms for the RD problem, particularly evident in classical distributions like discretized Gaussian and Laplacian sources \cite{wu2022communication,book_element}. 
Furthermore, the CBA algorithm also showcases remarkable efficiency in addressing the DR problem. 
Additionally, experiments conducted on sources with discrete isolated reproductions such as the uniform source \cite[Sec. VI]{rose94} and sources with bifurcation points in the RD curves such as those in \cite[Example 2.7.3]{berger1971}\cite{agmon2022root}, confirm the stability and efficiency of the CBA algorithm in contrast to the original BA algorithm. 
In these cases, the slope of RD curves undergoes significant variations across distortion values, and even slight perturbations in the value of slope can result in substantial errors when computing the RD function for a given distortion. 
Therefore, the accurate determination of the slope $\lambda$ corresponding to a particular distortion $D$ is crucial for precise computation of the RD function, and this task is handled with great efficiency by the CBA algorithm via Newton's method.

The remaining part of this paper is organized as follows. 
The derivation of the CBA algorithm for RD and DR functions is given in Section \ref{Sec_2_CBA}, and its {convergence and computational complexity are analyzed} in Section \ref{Sec_3_converge}. 
The numerical experiments are presented in Section \ref{Sec_4_numerical}. 
Finally Section \ref{Sec_5_conclusion} concludes this work.

\section{Constrained BA Algorithm} \label{Sec_2_CBA}
Consider a discrete memoryless source $X \in \mcX$ with reproduction $Y \in \mcY$, where $\mcX=\{x_1,\cdots,x_M\}$ and $\mcY=\{y_1,\cdots,y_N\}$ are finite alphabets. 
In the sequel, denote $p_{i}=P_{X}(x_{i})$, $r_{j}=P_{Y}(y_{j})$, $w_{ij}=P_{Y| X}(y_{j}| x_{i})$, and $d_{ij}=d(x_{i},y_{j})$ for simplicity. 
As a well known property (see, e.g., \cite[Thm. 9.18 and Cor. 9.19]{book_BA}), the RD function $R(D)$ is infinity for $D < D_{\min} \triangleq \min_{i, j} d_{ij}$ and zero for $D \geq D_{\max} \triangleq \min_{j} \sum_{i} p_i d_{ij}$, and furthermore, for $D \in [D_{\min}, D_{\max}]$, the expected distortion constraint in \eqref{rate_dis_tor_def} is active, i.e., attaining equality.
Therefore, we only need to focus on the case of $D \in [D_{\min}, D_{\max})$ in the following analysis.
In this case, the RD function $R(D)$ can be formulated as the following constrained optimization problem \cite{book_element,berger1971,wu2022communication},  
\begin{subequations} \label{semi_CommOT_model}
\begin{align}
&\mmin_{\bdw, \bdr} & {f_{R}(\bdw,\bdr) \triangleq} \sum_{i=1}^{M} \sum_{j=1}^{N} w_{i j}p_{i} \left[\log w_{i j}-\log r_{j}\right] \label{semi_CommOT_model_a} \\
&\text{subject to} & \sum_{i=1}^{M}\sum_{j=1}^{N} w_{i j} p_{i} d_{ij} \leq D, \quad \sum_{j=1}^{N} w_{i j}=1, ~ \forall i, \label{semi_CommOT_model_b} \\
& & \sum_{i=1}^{M} w_{ij}p_{i} = r_{j}, ~ \forall j, \quad\quad \sum_{j=1}^{N} r_{j}=1. \label{semi_CommOT_model_c}
\end{align}
\end{subequations}

On the other hand, we also consider the DR function $D(R)$ \cite[Chap. 10, Sec. 2]{book_element}, which is known as the inverse of the RD function and widely applied for quantization problems \cite{gray1998quantization} in source coding and lossy compression, written as the following form, 
\begin{subequations} \label{DR_model}
\begin{align}
&\mmin_{\bdw, \bdr} & {f_{D}(\bdw) \triangleq} \sum_{i=1}^{M} \sum_{j=1}^{N} w_{i j}p_{i} d_{ij} \label{DR_model_a} \\
&\text{subject to} & \sum_{i=1}^{M} \sum_{j=1}^{N} w_{i j}p_{i} \left[\log w_{i j}-\log r_{j}\right]\leq R, \quad \sum_{j=1}^{N} w_{i j}=1, ~ \forall i, \label{DR_model_b} \\
& &\sum_{i=1}^{M} w_{ij}p_{i} = r_{j}, ~ \forall j, \quad\quad \sum_{j=1}^{N} r_{j}=1. \label{DR_model_c}
\end{align}
\end{subequations}
\begin{remark} \label{remark_1}
In the models described above, we introduce slackness variables $r_{j}$ and constraints on marginal distributions $\sum_{i=1}^{M} w_{ij}p_{i} = r_{j}, \forall j$, inspired by our recent work \cite{wu2022communication}.
However, adhering to the procedure outlined in the seminal work \cite{blahut1972computation}, we can prove that the constraints $\sum_{i=1}^{M} w_{ij}p_{i} = r_{j}, \forall j$ can be relaxed in the subsequent Lagrangian analysis. This is achieved by considering an alternative optimization approach in the $r_{j}$ direction, which is thoroughly detailed in the following subsections.
This seemingly straightforward yet crucial relaxation not only simplifies the problem formulation during the Lagrangian analysis in this work but also yields a closed-form solution in the $r_{j}$ direction, echoing the insights from \cite{blahut1972computation}.
As shown in \cite{wu2022communication}, the constraints $\sum_{i=1}^{M} w_{ij}p_{i} = r_{j}, \forall j$ can be regarded as marginal distribution constraints in an entropy regularized optimal transport (OT) model, and hence the relaxation of these constraints can be viewed as the relaxation of one-side marginal distribution constraints and the relaxed problem can be further viewed as a semi-relaxed entropy regularized OT model \cite{peyre2019computational}.
\end{remark}

Directly, by introducing the Lagrange multipliers $\eta\in\mbbR$, $\bda\in\mbbR^{M}$, and $\lambda\in\mbbR^{+}$, and by removing the marginal distribution constraints in light of Remark \ref{remark_1} above, the Lagrangian of the constrained optimization problem \eqref{semi_CommOT_model}, is written as: 
\begin{equation}\label{Lagrangian1}
\begin{aligned}
\mathcal{L}_{R}(\bdw, \bdr; {\eta}, \bda, {\lambda})&=\sum_{i=1}^{M}\sum_{j=1}^{N} w_{i j}p_{i} \left[ \log w_{ij}-\log r_{j}\right]+{\eta}\left(\sum_{j=1}^{N} r_{j}-1\right) \\
&+\sum_{i=1}^{M} \alpha_{i} \left(\sum_{j=1}^{N} w_{ij}-1\right)+{\lambda}\left(\sum_{i=1}^{M}\sum_{j=1}^{N} w_{i j} p_{i} d_{ij}-D\right).
\end{aligned}
\end{equation}
Similarly, by introducing multipliers $\eta\in\mbbR$, $\bda\in\mbbR^{M}$, and $\zeta\in\mbbR^{+}$, the Lagrangian of \eqref{DR_model} is written as: 
\begin{equation}\label{DR_Lagrangian}
\begin{aligned}
\mathcal{L}_{D}(\bdw, \bdr; {\eta}, \bda, {\zeta})&=\zeta\left(\sum_{i=1}^{M}\sum_{j=1}^{N} w_{i j}p_{i} \left[ \log w_{ij}-\log r_{j}\right]-R\right)+{\eta}\left(\sum_{j=1}^{N} r_{j}-1\right) \\
&+\sum_{i=1}^{M} \alpha_{i} \left(\sum_{j=1}^{N} w_{ij}-1\right)+\sum_{i=1}^{M}\sum_{j=1}^{N} w_{i j} p_{i} d_{ij}.
\end{aligned}
\end{equation}
In the following proposed algorithms, we adopt a different approach from the BA algorithm by updating the multipliers in each iteration, rather than keeping them fixed.
This difference essentially stems from viewing the RD and DR functions as constrained optimization problems, rather than unconstrained RD and DR Lagrangian with fixed multipliers, and thus, we term our algorithm as the \textbf{C}onstrained \textbf{B}lahut-\textbf{A}rimoto (\textbf{CBA}) algorithm. 
Specifically, the distortion constraint in \eqref{semi_CommOT_model_b} and the rate constraint in \eqref{DR_model_b}, involved into the objective functions of RD and DR Lagrangian by fixing the multipliers in the BA algorithm, are now treated as an additional constraint in our algorithm. 
Notably, the distortion constraint and the rate constraint can be converted into a root-finding problem with respect to a monotonic function. 
Thus, they can be efficiently handled by a one-dimensional Newton's method, which brings visibly advantage on efficiency. 
In the sequel, we give the details of the CBA algorithm for the RD and DR functions.

\subsection{Constrained BA algorithm for the RD function}\label{subsec_2_RD}
The main steps for computing the RD function in each direction are presented as follows, and a pseudo-code is listed in Algorithm \ref{alg:OT_RD}.
\textbf{
\begin{itemize}
    \item[(i)] Fix $\bdr$, and obtain the multiplier $\lambda\in\mbbR^{+}$ by applying Newton's method to find the unique root of the following monotonic univariate function $G_{R}(\lambda)$:
    \begin{equation}\label{def_G_R}
    G_{R}(\lambda) \triangleq \sum_{i=1}^{M}\sum_{j=1}^{N} d_{ij} p_{i} \left(e^{-\lambda d_{ij}}r_{j}\bigg/\left(\sum_{j=1}^{N} e^{-\lambda d_{ij}}r_{j}\right)\right)-D.
    \end{equation}
    Update the primal variables $w_{ij}$ in closed-form with $\lambda$ obtained in \eqref{def_G_R}:
    \begin{equation}\label{eq_w}
        w_{ij}=e^{-\lambda d_{ij}}r_{j}\bigg/\left(\sum_{j=1}^{N} e^{-\lambda d_{ij}}r_{j}\right). 
    \end{equation}
    \item[(ii)] 
    Fix $\bdw$, and update the primal variables $r_{j}$ as follows:
    \begin{equation}\label{eq_r}
        r_{j} = \sum_{i=1}^{M} w_{i j}p_{i}.
    \end{equation}
\end{itemize}
}
\begin{algorithm}[H]
	\caption {Constrained Blahut-Arimoto (CBA) algorithm for RD function}
	\label{alg:OT_RD}
	\begin{algorithmic}[1]
		\Require Distortion measure $d_{ij}$, marginal distribution $p_{i}$, iteration number $iter$.
		\State \textbf{Initialization:} $r_{j}=1/N$
		\For{$\ell = 1 : iter$}
		\State Solve $G_{R}(\lambda) = 0$ in \eqref{def_G_R} for $\lambda\in\mbbR^{+}$ with Newton's method
		\State Update $w_{ij} \gets e^{-\lambda d_{ij}}r_{j}\Big/\left(\sum_{j=1}^{N} e^{-\lambda d_{ij}}r_{j}\right)$
        \State Update $r_{j} \gets \sum_{i=1}^{M} w_{i j}p_{i}$ 
		\EndFor
		\State \textbf{end}\\
		\Return $\sum_{i=1}^{M} \sum_{j=1}^{N} (w_{i j}p_{i}) \left[\log w_{i j}\!\!-\!\!\log r_{j}\right]$
	\end{algorithmic}
\end{algorithm}
In the following, we will provide a detailed derivation of the CBA algorithm for the RD function.
Specifically, we will iteratively update the variables $\bdw$ and $\bdr$ by solving the constrained optimization problem \eqref{semi_CommOT_model}.
\textbf{First, fix the variable $\bdr$, and minimize \eqref{semi_CommOT_model} as an optimization problem with respect to the variable $\bdw$ only.}
Taking the partial derivative of $\mathcal{L}_{R}\left(\bdw, \bdr; \eta, \bda, \lambda\right)$ with respect to the primal variable $\bdw$, we obtain the following first-order condition:
\begin{equation*} \label{dW}
\frac{\partial \mathcal{L}_{R}}{\partial w_{ij}}=p_{i}\left(1+\log w_{i j}-\log r_{j}\right)+\alpha_{i}+\lambda p_{i} d_{ij} = 0,
\end{equation*}
and it further yields the representation of $\bdw$ by the multipliers:
\begin{equation} \label{update_W}
w_{ij}=e^{-\frac{\alpha_{i}}{p_{i}}-1}e^{-\lambda d_{i j}}r_{j}.
\end{equation}
Then, we substitute \eqref{update_W} into the two constraints in \eqref{semi_CommOT_model_b} respectively, thus obtaining
\begin{equation} \label{up_alpha}
e^{-\frac{\alpha_{i}}{p_{i}}-1}\left(\sum_{j=1}^{N} e^{-\lambda d_{ij}} r_{j}\right)=1, \quad i=1, \cdots, M,
\end{equation}
\begin{equation} \label{up_lambda}
\sum_{i=1}^{M}\sum_{j=1}^{N} e^{-\frac{\alpha_{i}}{p_{i}}-1}e^{-\lambda d_{ij}}r_{j} p_{i} d_{ij}=D.
\end{equation}
By combining equations \eqref{up_alpha} and \eqref{up_lambda} and eliminating $\bda$, we obtain the update rule for the multiplier $\lambda$, i.e., 
\begin{equation*} \label{root_finding_problem}
G_{R}(\lambda) \triangleq \sum_{i=1}^{M}\sum_{j=1}^{N} d_{ij} p_{i} \left(e^{-\lambda d_{ij}}r_{j}\bigg/\left(\sum_{j=1}^{N} e^{-\lambda d_{ij}}r_{j}\right)\right)-D=0.
\end{equation*}
Since we focus on $D \in [D_{\min}, D_{\max})$, we have
\begin{equation*}
    G_{R}(0)=\sum_{i,j} p_i d_{ij}r_j-D \geq D_{\max} - D > 0.
\end{equation*}
On the other hand, as $\lambda\rightarrow\infty$, we have $G_{R}(\lambda)\rightarrow\sum_i p_i\min_j d_{ij} - D \leq D_{\min} - D \leq 0$. 
By the intermediate value theorem, $G_{R}(\lambda)$ has at least one root $\lambda\in\mbbR^{+}$.
Furthermore, since the derivative of the univariate function $G_{R}(\lambda)$ is 
\begin{equation*}
G_{R}^{\prime}(\lambda)=-\sum_{i=1}^{M} p_{i}\left[\frac{\left(\sum_{j=1}^{N}d_{ij}^{2}e^{-\lambda d_{ij}}r_{j}\right)\left(\sum_{j=1}^{N}e^{-\lambda d_{ij}}r_{j}\right)-\left(\sum_{j=1}^{N}d_{ij}e^{-\lambda d_{ij}}r_{j}\right)^{2}}{\left(\sum_{j=1}^{N} e^{-\lambda d_{ij}}r_{j}\right)^{2}}\right]\leq 0,
\end{equation*}
where the inequality follows from the  Cauchy-Schwarz inequality, $G_{R}(\lambda)$ is monotonic, and hence has a unique root $\lambda\in\mbbR^{+}$, which can be obtained via a one-dimensional root-finding step by Newton's method, typically with only a few iterations due to its second order convergence property.
By combining \eqref{update_W} and \eqref{up_alpha}, we can obtain the optimal $\bdw$, denoted as $\Tilde{\bdw}$, under fixed $\bdr$, after updating $\lambda$ via finding the root of the monotonic univariate function $G_{R}(\lambda)$: 
\begin{equation*}
\Tilde{w}_{ij}=e^{-\lambda d_{ij}}r_{j}\bigg/\left(\sum_{j=1}^{N} e^{-\lambda d_{ij}}r_{j}\right), \quad i = 1, \cdots, M,\quad j = 1, \cdots, N.
\end{equation*}

\textbf{Second, fix the variable $\bdw$, and minimize \eqref{semi_CommOT_model} as an optimization problem with respect to the variable $\bdr$ only. }
Similarly, taking the partial derivative of $\mathcal{L}_{R}\left(\bdw, \bdr; \eta, \bda, \lambda\right)$ with respect to the primal variable $\bdr$, we obtain the first-order condition:
\begin{equation*}
\frac{\partial \mathcal{L}_{R}}{\partial r_{j}}=-\sum_{i=1}^{M} w_{i j}p_{i} \frac{1}{r_{j}}+\eta=0,
\end{equation*}
and it further yields the representation of $\bdr$ by the multiplier $\eta$:
\begin{equation}\label{from_of_r}
r_{j} = \left(\sum_{i=1}^{M} w_{i j}p_{i} \right)\Big/\eta.
\end{equation}
Substituting \eqref{from_of_r} into the equality constraint in \eqref{semi_CommOT_model_c}, we have that the multiplier $\eta$ should be updated to satisfy the following equation: 
\begin{equation*}
F_{R}(\eta) \triangleq \sum_{j=1}^{N}\left[\left(\sum_{i=1}^{M} w_{i j}p_{i} \right)\Big/\eta\right]-1=0. \label{F_def}
\end{equation*}
Here, $F_{R}(\eta)$ is a monotonic function with a unique real root $\eta=1$, due to the fact 
\begin{equation*}
\sum_{i=1}^{M}\sum_{j=1}^{N} w_{ij}p_{i}=1.
\end{equation*}
Hence, we obtain the optimal $\bdr$, denoted $\Tilde{\bdr}$, under fixed $\bdw$, as
\begin{equation*}
\Tilde{r}_j=\sum_{i=1}^{M} w_{i j}p_{i}, \quad j = 1, \cdots, N.
\end{equation*}

\subsection{Constrained BA algorithm for DR function}\label{subsec_2_DR}

For the DR problem, a fundamental distinction arises. Unlike the RD function, the objective function in \eqref{DR_model} cannot be written in the form of alternating projections on two convex sets. 
Consequently, existing algorithms \cite{hayashi2023bregman,wu2022communication} may encounter challenges in this scenario.
%
%
However, as we will demonstrate, our proposed Lagrangian based approach remains effective and successful in addressing the DR problem.
In the following, the main steps and the pseudo-code for the DR function are presented.

\textbf{
\begin{itemize}
    \item[(i)] Fix $\bdr$, and obtain $\lambda=1/\zeta\in\mbbR^{+}$ by applying Newton's method to find the unique root of the following monotonic univariate function:
    \begin{equation}\label{DR_def_G_D}
    G_{D}(\lambda) \triangleq -\sum_{i=1}^M p_i \log\left(\sum_{j=1}^N r_j e^{-\lambda d_{ij}}\right) -\lambda\sum_{i=1}^{M}\sum_{j=1}^{N} d_{ij} p_{i} \left(e^{-\lambda d_{ij}}r_{j}\bigg/\left(\sum_{j=1}^{N} e^{-\lambda d_{ij}}r_{j}\right)\right)-R=0.
    \end{equation}
     Update the primal variables $w_{ij}$ in closed-form with $\lambda$ obtained in \eqref{DR_def_G_D}:
    \begin{equation}\label{DR_eq_w}
        w_{ij}=e^{-\lambda d_{ij}}r_{j}\bigg/\left(\sum_{j=1}^{N} e^{-\lambda d_{ij}}r_{j}\right). 
    \end{equation}
    \item[(ii)] 
    Fix $\bdw$, and update the primal variables $r_{j}$ as follows:
    \begin{equation}\label{DR_eq_r}
        r_{j} = \sum_{i=1}^{M} w_{i j}p_{i}.
    \end{equation}
\end{itemize}
}

\begin{algorithm}[H]
	\caption {Constrained Blahut-Arimoto (CBA) algorithm for DR function}
	\label{alg:OT_DR}
	\begin{algorithmic}[1]
		\Require Distortion measure $d_{ij}$, marginal distribution $p_{i}$, iteration number $iter$.
		\State \textbf{Initialization:} $r_{j}=1/N$
		\For{$\ell = 1 : iter$}
		\State Solve $G_{D}(\lambda) = 0$ in \eqref{DR_def_G_D} for $\lambda\in\mbbR^{+}$ with Newton's method
		\State Update $w_{ij} \gets e^{-\lambda d_{ij}}r_{j}\Big/\left(\sum_{j=1}^{N} e^{-\lambda d_{ij}}r_{j}\right)$
        \State Update $r_{j}\gets \sum_{i=1}^{M} w_{i j}p_{i}$ 
		\EndFor
		\State \textbf{end}\\
		\Return $\sum_{i=1}^{M} w_{i j}p_{i} d_{ij}$
	\end{algorithmic}
\end{algorithm}
In the following, we will provide a detailed derivation of the CBA algorithm for the DR function. 
Specifically, we will iteratively update the variables $\bdw$ and $\bdr$ by solving the constrained optimization problem \eqref{DR_model}. 
\vspace{+.3in}

\textbf{First, fix the variable $\bdr$, and update $\bdw$ and the corresponding dual variables.}
Taking the partial derivative of $\mathcal{L}_{D}\left(\bdw, \bdr; \eta, \bda, \zeta\right)$ with respect to the primal variable $\bdw$, we obtain the first-order condition:
\begin{equation*}
\frac{\partial \mathcal{L}_{D}}{\partial w_{ij}}=\zeta p_{i}\left(1+\log w_{i j}-\log r_{j}\right)+\alpha_{i}+ p_{i} d_{ij} = 0,
\end{equation*}
and it further yields the representation of $\bdw$ by the multipliers:
\begin{equation} \label{DR_update_W}
w_{ij}=e^{-\frac{\alpha_{i}}{\zeta p_{i}}-1}e^{-1/\zeta d_{i j}}r_{j}.
\end{equation}
Then, we substitute \eqref{DR_update_W} into the two constraints in \eqref{DR_model_b} respectively, thus obtaining
\begin{equation} \label{DR_up_alpha}
e^{-\frac{\alpha_{i}}{\zeta p_{i}}-1}\left(\sum_{j=1}^{N} e^{-1/\zeta d_{ij}} r_{j}\right)=1, \quad i=1, \cdots, M,
\end{equation}
\begin{equation} \label{DR_up_lambda}
\sum_{i=1}^{M}\sum_{j=1}^{N} e^{-\frac{\alpha_{i}}{\zeta p_{i}}-1}e^{-1/\zeta d_{i j}}r_{j}p_{i} \left[-\frac{\alpha_{i}}{\zeta p_{i}}-1-\frac{1}{\zeta d_{i j}} \right]=R.
\end{equation}
By combining equations \eqref{DR_up_alpha} and \eqref{DR_up_lambda} and eliminating $\bda$, and then setting $\lambda=1/\zeta$, we obtain the update rule for the Lagrange multiplier $\lambda$ as follows:
\begin{equation*}
G_{D}(\lambda) \triangleq -\sum_{i=1}^M p_i \log\left(\sum_{j=1}^N r_j e^{-\lambda d_{ij}}\right) -\lambda\sum_{i=1}^{M}\sum_{j=1}^{N} d_{ij} p_{i} \left(e^{-\lambda d_{ij}}r_{j}\bigg/\left(\sum_{j=1}^{N} e^{-\lambda d_{ij}}r_{j}\right)\right)-R=0.
\end{equation*}
Since the derivative of the univariate function $G_{D}(\lambda)$ is 
\begin{equation*}
G_{D}^{\prime}(\lambda)=\lambda\sum_{i=1}^{M} p_{i}\left[\frac{\left(\sum_{j=1}^{N}d_{ij}^{2}e^{-\lambda d_{ij}}r_{j}\right)\left(\sum_{j=1}^{N}e^{-\lambda d_{ij}}r_{j}\right)-\left(\sum_{j=1}^{N}d_{ij}e^{-\lambda d_{ij}}r_{j}\right)^{2}}{\left(\sum_{j=1}^{N} e^{-\lambda d_{ij}}r_{j}\right)^{2}}\right]\geq 0,
\end{equation*}
where the inequality follows from the Cauchy-Schwarz inequality, $G_{D}(\lambda)$ is monotonic, and hence has a unique root $\lambda\in\mathbb{R}^{+}$, which can be efficiently found using Newton's method in a one-dimensional root-finding step, typically converging in just a few iterations due to its second-order convergence property. 
By combining \eqref{DR_update_W} and \eqref{DR_up_alpha}, we obtain the update rule of $\bdw$ after updating the multiplier $\lambda$ via finding the root of the monotonic univariate function $G_{D}(\lambda)$:
\begin{equation*}
{w}_{ij}=e^{-\lambda d_{ij}}r_{j}\bigg/\left(\sum_{j=1}^{N} e^{-\lambda d_{ij}}r_{j}\right), \quad i = 1, \cdots, M,\quad j = 1, \cdots, N.
\end{equation*}

\textbf{Second, fix the variable $\bdw$, and update $\bdr$ and the corresponding dual variables.}
Similarly, taking the partial derivative of $\mathcal{L}_{D}\left(\bdw, \bdr; \eta, \bda, \zeta\right)$ with respect to the primal variable $\bdr$, we obtain the first-order condition:
\begin{equation*}
\frac{\partial \mathcal{L}_{D}}{\partial r_{j}}=-\zeta\sum_{i=1}^{M} w_{i j}p_{i} \frac{1}{r_{j}}+\eta=0,
\end{equation*}
and it further yields the representation of $\bdr$ by the multiplier $\eta$:
\begin{equation}\label{DR_from_of_r}
r_{j} = \zeta\left(\sum_{i=1}^{M} w_{i j}p_{i} \right)\bigg/\eta.
\end{equation}
Substituting \eqref{from_of_r} into the equality constraint in \eqref{semi_CommOT_model_c}, we note that the multiplier $\eta$ should be updated to satisfy the following equation:
\begin{equation*}
F_{D}(\eta) \triangleq \sum_{j=1}^{N}\zeta\left[\left(\sum_{i=1}^{M} w_{i j}p_{i} \right)\bigg/\eta\right]-1=0. \label{DR_F_def}
\end{equation*}
Here, $F_{D}(\eta)$ is a monotonic function with a unique real root $\eta=\zeta$, due to the fact 
\begin{equation*}
\sum_{i=1}^{M}\sum_{j=1}^{N} w_{ij}p_{i}=1.
\end{equation*}
Hence, we obtain the update of $\bdr$, under fixed $\bdw$, as
\begin{equation*}
{r}_j=\sum_{i=1}^{M} w_{i j}p_{i}, \quad j = 1, \cdots, N.
\end{equation*}

\bigskip\bigskip To this end, we have developed the CBA algorithm for RD and DR functions. This constitutes a unified algorithm framework for the above functions. In fact, optimization formulations of the RD function \eqref{semi_CommOT_model} and the DR function \eqref{DR_model} are similar. These two functions can be regarded as inverse functions of each other. Thus, we believe that the CBA algorithm is highly consistent with the characteristics of the above problems and is the most appropriate choice.

\section{Convergence Analysis} \label{Sec_3_converge}
In this section, we prove the convergence of the proposed CBA algorithm for the RD and DR functions. 
Specifically, we provide a convergence proof for the proposed CBA algorithm, demonstrating that it converges to the solution of the RD and DR functions, with a convergence rate of $O(1/n)$. 
Moreover, we establish that the CBA algorithm is capable of providing an approximate solution close to the optimal solution within an arbitrary $\varepsilon>0$, with arithmetic operations of $O\left(\frac{MN\log N}{\varepsilon}(1+\log |\log \varepsilon|)\right)$. 

The main results are summarized as follows.
\begin{theorem}\label{thm: 1}
Denoting by $(\bdw^{(n)},\bdr^{(n)})$ the sequence generated by the CBA algorithm and by $(\bdw^{*},\bdr^{*})$ the global optimizer of RD function \eqref{semi_CommOT_model}, we have 
\begin{equation*}
f_{R}(\bdw^{(n)},\bdr^{(n)})-f_{R}(\bdw^{*},\bdr^{*})\simeq O(1/n),
\end{equation*}
i.e., the CBA algorithm converges to the RD function with rate $O(1/n)$.
\end{theorem}
\begin{theorem} \label{thm: 3}
The CBA algorithm for RD function outputs solution $(\bdw^{(n)}, \bdr^{(n)})$ satisfying
\begin{equation}  
    f_{R}(\bdw^{(n)}, \bdr^{(n)})-f_{R}(\bdw^{*}, \bdr^{*}) \leq \varepsilon
\end{equation}
with $O\left(\frac{MN\log N}{\varepsilon}(1+\log|\log \varepsilon|)\right)$ arithmetic operations. 
\end{theorem}
\begin{remark}
The algorithm in \cite{hayashi2023bregman} has an estimation of $O\left(\frac{MN\log N}{\varepsilon}(\log N+|\log \varepsilon|)\right)$ arithmetic operations. 
Our proposed CBA algorithm exploits the monotonic property of the function $G_{R}(\lambda)$ and uses the Newton's method instead of the binary search, which leads to significantly more efficient and robust results, as demonstrated in both theoretical and numerical experiments. 
\end{remark}
\begin{theorem}\label{thm: 2}
Denoting by $(\bdw^{(n)},\bdr^{(n)})$ the sequence generated by the CBA algorithm, and by $(\bdw^{*},\bdr^{*})$ the global optimizer of DR function \eqref{DR_model}, we have 
\begin{equation*}
f_{D}(\bdw^{(n)})-f_{D}(\bdw^{*})\simeq O(1/n),
\end{equation*}
i.e., the CBA algorithm converges to the DR function with rate $O(1/n)$.
\end{theorem}
\begin{theorem} \label{thm: 4}
The CBA algorithm for DR function outputs solution $(\bdw^{(n)}, \bdr^{(n)})$ satisfying
\begin{equation}  
    f_{D}(\bdw^{(n)})-f_{D}(\bdw^{*}) \leq \varepsilon
\end{equation}
with $O\left(\frac{MN\log N}{\varepsilon}(1+\log|\log \varepsilon|)\right)$ arithmetic operations. 
\end{theorem}
In the following subsections, we provide detailed proofs of the above theorems.
\subsection{Convergence analysis for the RD function} \label{subsec_3_RD}
Before the proof of Theorem \ref{thm: 1}, we need some lemmas. 
First, we estimate the reduction of the objective function $f_{R}(\bdw,\bdr)$ with respect to $\bdw$, for a fixed $\bdr$.
Here, we denote by $D_{KL}(\cdot\|\cdot)$ the Kullback-Leibler (KL) divergence and by $\bdw_{i}$ the $i_{th}$ row of $\bdw$.

\begin{lemma}\label{lem: 1}
For a fixed $\bdr$, the choice of $\bdw$, denoted by $\Tilde{\bdw}(\bdr)$, that minimizes the objective function $f_{R}(\bdw,\bdr)$ under the constraint \eqref{semi_CommOT_model_b}, is given by
\begin{equation*}
\Tilde{w}_{ij}=e^{-\lambda d_{ij}}r_{j}\bigg/\left(\sum_{k=1}^{N} e^{-\lambda d_{ik}}r_{k}\right),
\end{equation*}
where $\lambda \in \mbbR^{+}$ is the unique root of the monotonic function $G_{R}(\lambda)$.
Further, for any other choice of $\bdw$ satisfying $\sum_{i,j}w_{ij}p_{i}d_{ij}=D$, the following inequality holds:
\begin{equation*}
f_{R}(\bdw,\bdr)-f_{R}(\Tilde{\bdw}(\bdr),\bdr) \geq D_{KL}\left(\sum_{i=1}^{M}p_{i}\bdw_{i}\Big\|\sum_{i=1}^{M}p_{i}\Tilde{\bdw}_{i}\right).
\end{equation*}
\end{lemma}
\begin{proofs} 
For a fixed $\bdr$, we have already obtained the optimal solution of the form $\Tilde{\bdw}(\bdr)$ in Section \ref{subsec_2_RD}.
Therefore, we only need to estimate a bound of the difference $f_{R}(\bdw,\bdr)-f_{R}(\Tilde{\bdw}(\bdr),\bdr)$.
Let $\lambda$ be the root of the monotonic univariate function $G_{R}(\lambda)$ with the fixed $\bdr$, as defined in \eqref{def_G_R}, i.e., $\lambda$ satisfying:
\begin{equation*} 
\sum_{i=1}^{M}\sum_{j=1}^{N} d_{ij} p_{i} \left(e^{-\lambda d_{ij}}r_{j}\bigg/\left(\sum_{k=1}^{N} e^{-\lambda d_{ik}}r_{k}\right)\right)-D=0.
\end{equation*}
For an arbitrary $\bdw$ satisfying $\sum_{i,j}w_{ij}p_{i}d_{ij}=D$, we have
\begin{multline*}
f_{R}\left(\bdw, \bdr\right)-f_{R}\left(\Tilde{\bdw}(\bdr), \bdr\right) = f_{R}\left(\bdw, \bdr\right)+\lambda D-f_{R}\left(\Tilde{\bdw}, \bdr\right)-\lambda D \\
=\left(f_{R}\left(\bdw, \bdr\right)\!+\!\lambda\sum_{i,j}p_{i}w_{ij}d_{ij}\right)\!-\!\left(f_{R}\left(\Tilde{\bdw}, \bdr\right)\!+\!\!\lambda\sum_{i,j}p_{i}\Tilde{w}_{ij}d_{ij}\right) \\
=\sum_{i,j}p_{i}w_{ij}\log\frac{w_{ij}}{r_{j}e^{-\lambda d_{ij}}}-\sum_{i,j}p_{i}\Tilde{w}_{ij}\left[\log\frac{e^{-\lambda d_{ij}}}{\sum_{k}r_{k}e^{-\lambda d_{ik}}} + \lambda d_{ij}\right] \\
=\sum_{i,j}p_{i}w_{ij}\log\frac{w_{ij}}{r_{j}e^{-\lambda d_{ij}}}-\sum_{i}\left[\sum_{j}p_{i}\Tilde{w}_{ij}\right]\log\frac{1}{\sum_{k}r_{k}e^{-\lambda d_{ik}}} \\
=\sum_{i,j}p_{i}w_{ij}\log\frac{w_{ij}}{r_{j}e^{-\lambda d_{ij}}}-\sum_{i}\left[\sum_{j}p_{i}{w}_{ij}\right]\log\frac{1}{\sum_{k}r_{k}e^{-\lambda d_{ik}}} \\
=\sum_{i,j}p_{i}w_{ij}\log\frac{w_{ij}\big(\sum_{k}r_{k}e^{-\lambda d_{ik}}\big)}{r_{j}e^{-\lambda d_{ij}}}\geq D_{KL}\left(\sum_{i=1}^{M}p_{i}\bdw_{i}\Big\|\sum_{i=1}^{M}p_{i}\Tilde{\bdw}_{i}\right).
\end{multline*}
Here, the last inequality is due to the convex property of KL divergence.
\end{proofs}
Next, we estimate the reduction of the objective function $f_{R}(\bdw,\bdr)$ with respect to $\bdr$, for a fixed $\bdw$.
\begin{lemma} \label{lem: 2}
For a fixed $\bdw$, the choice of $\bdr$, denoted by $\Tilde{\bdr}(\bdw)$, that minimizes the objective function $f_{R}(\bdw,\bdr)$ under the constraint \eqref{semi_CommOT_model_c}, is given by
\begin{equation*}
\tilde{r}_{j}=\sum_{i=1}^{M}w_{ij}p_{i}.
\end{equation*}
For any other choice of $\bdr$, the following equality holds:
\begin{equation*}
f_{R}(\bdw,\bdr)-f_{R}(\bdw,\Tilde{\bdr}(\bdw))=D_{KL}(\Tilde{\bdr}(\bdw)\|\bdr).
\end{equation*}
\end{lemma}
\begin{proofs}
For a fixed $\bdw$, we have already obtained the optimal solution $\Tilde{\bdr}(\bdw)$ in Section \ref{subsec_2_RD}. 
Therefore, our focus is to study the difference $f_{R}(\bdw,\bdr)-f_{R}(\bdw,\Tilde{\bdr}(\bdw))$.

For an arbitrary $\bdr$, we have
\begin{multline*}
    f_{R}(\bdw,\bdr)-f_{R}(\bdw,\Tilde{\bdr}(\bdw))=\sum_{i,j} w_{i j}p_{i} \left[\log w_{i j}-\log r_{j}\right]-\sum_{i,j} w_{i j}p_{i} \left[\log w_{i j}-\log \Tilde{r}_{j}\right]\\
    =\sum_{i,j} w_{i j}p_i(\log\Tilde{r}_{j}-\log r_j)=\sum_{j}\Tilde{r}_j\log\frac{\Tilde{r}_{j}}{r_j}=D_{KL}(\Tilde{\bdr}(\bdw)\|\bdr).
\end{multline*}
\end{proofs}
By applying the above two lemmas, we can establish the following non-increasing property of the objective function $f_{R}(\bdw,\bdr)$ during iterations.
\begin{lemma} \label{lem: 3}
Letting $(\bdw^{(n)},\bdr^{(n)})$ be the sequence generated by the CBA algorithm where $\bdw^{(n)}=\Tilde{\bdw}(\bdr^{(n)})$ and $\bdr^{(n)}=\Tilde{\bdr}(\bdw^{(n-1)})$, we have the following property:
\begin{equation*}
f_{R}(\bdw^{(n)},\bdr^{(n)})\leq f_{R}(\bdw^{(n-1)},\bdr^{(n-1)}).
\end{equation*}
\end{lemma}
\begin{proofs} 
Applying Lemmas \ref{lem: 1} and \ref{lem: 2}, and noticing $\bdw^{(n)}=\Tilde{\bdw}(\bdr^{(n)})$ and $\bdr^{(n)}=\Tilde{\bdr}(\bdw^{(n-1)})$, we have
\begin{multline*}
f_{R}(\bdw^{(n)},\bdr^{(n)})-f_{R}(\bdw^{(n-1)},\bdr^{(n-1)}) \\
=\left(f_{R}(\bdw^{(n)},\bdr^{(n)}) - f_{R}(\bdw^{(n-1)},\bdr^{(n)})\right)+\left(f_{R}(\bdw^{(n-1)},\bdr^{(n)}) - f_{R}(\bdw^{(n-1)},\bdr^{(n-1)})\right) \\
\leq -D_{KL}\left(\sum_{i} p_{i}\bdw^{(n-1)}_{i}\Big\|\sum_{i} p_{i} \bdw^{(n)}_{i}\right)-D_{KL}(\bdr^{(n)}\|\bdr^{(n-1)})\leq 0.
\end{multline*}
Thus, we have $f_{R}(\bdw^{(n)},\bdr^{(n)})\leq f_{R}(\bdw^{(n-1)},\bdr^{(n-1)})$.
\end{proofs}
At this point, we are ready to give the proof of Theorem \ref{thm: 1}.

\begin{proofs}
Applying Lemmas \ref{lem: 1} and \ref{lem: 2}, and noticing $\bdw^{(n)}=\Tilde{\bdw}(\bdr^{(n)})$ and $\ \bdr^{*}=\Tilde{\bdr}(\bdw^{*})$, we have 
\begin{multline*}
f_{R}(\bdw^{(n)},\bdr^{(n)})-f_{R}(\bdw^{*},\bdr^{*})=\left(f_{R}(\bdw^{(n)},\bdr^{(n)})-f_{R}(\bdw^{*},\bdr^{(n)})\right)+\left(f_{R}(\bdw^{*},\bdr^{(n)})-f_{R}(\bdw^{*},\bdr^{*})\right) \\
\leq-D_{KL}\Big(\sum_{i}p_{i}\bdw_{i}^{*}\Big\|\sum_{i}p_{i}\bdw_{i}^{(n)}\Big)+ D_{KL}\Big(\bdr^{*}\big\|\bdr^{(n)}\Big)=-D_{KL}\left(\bdr^{*}\|\bdr^{(n+1)}\right)+D_{KL}\left(\bdr^{*}\|\bdr^{(n)}\right),
\end{multline*}
for $n = 1, 2, \cdots$.
Then, by summing up these inequalities, we have
\begin{equation*}
\sum_{k=1}^{n}\Big(f_{R}(\bdw^{(k)},\bdr^{(k)})-f_{R}(\bdw^{*},\bdr^{*})\Big)\leq D_{KL}\left(\bdr^{*}\|\bdr^{(1)}\right)-D_{KL}\left(\bdr^{*}\|\bdr^{(n+1)}\right)\leq D_{KL}\left(\bdr^{*}\|\bdr^{(1)}\right)\triangleq C,
\end{equation*}
where $C$ is a constant.
Since $f_{R}(\bdw^{(k)},\bdr^{(k)})\leq f_{R}(\bdw^{(k-1)},\bdr^{(k-1)})$ according to Lemma \ref{lem: 3}, we obtain
\begin{equation*}
    0\leq n\left( f_{R}(\bdw^{(n)},\bdr^{(n)})-f_{R}(\bdw^{*},\bdr^{*})\right)\leq \sum_{k=1}^{n}\Big(f_{R}(\bdw^{(k)},\bdr^{(k)})-f_{R}(\bdw^{*},\bdr^{*})\Big)\leq C,
\end{equation*}
i.e., $f_{R}(\bdw^{(n)},\bdr^{(n)})-f_{R}(\bdw^{*},\bdr^{*})\simeq O(1/n)$. 
\end{proofs}
\vspace{+.2in}
\begin{remark}
It has been shown in the literature (see, e.g., \cite[Chap. 8]{csiszar11}) that the BA algorithm also converges to the RD function at rate $O(1/n)$. But it is interesting to note the following difference:
\begin{equation} \label{converg_rate}
\begin{aligned}
& \text{BA:} &(R^{(n)}+\lambda D^{(n)})-(R_\lambda^{*}+\lambda D_\lambda^{*})\simeq O\left(1/n\right), \\
& \text{CBA:} & R^{(n)}-R_D^{*}\simeq O\left(1/n\right).
\end{aligned}
\end{equation}
Here, for the BA algorithm, $R^{(n)}=f_{R}(\bdw^{(n)},\bdr^{(n)})$ and $D^{(n)}=\sum_{i,j} w_{ij}^{(n)} p_i d_{ij}$ are the rate and distortion computed during iterations respectively, for a fixed $\lambda$, and $(R_\lambda^{*}, D_\lambda^{*})$ corresponds to the optimal solution that minimizes the RD Lagrangian under this fixed $\lambda$. 
In contrast, for the CBA algorithm, $R^{(n)}=f_{R}(\bdw^{(n)},\bdr^{(n)})$ is the rate computed subject to the given target distortion $D$ throughout, with consecutively updated $\lambda$, and $R_D^{*}$ is the RD function value for the given $D$. 
That is, the BA algorithm minimizes the RD Lagrangian for a given multiplier $\lambda$, whereas the CBA algorithm directly approaches the optimal rate of the RD function for a given $D$.
\end{remark}

Furthermore, we provide an analysis on the computational complexity of the proposed CBA algorithm for $\varepsilon$-approximation of the optimal solution.
In this regard, we present the proof of Theorem \ref{thm: 3}.

\begin{proofs}
Since the monotonic function $G_{R}(\lambda)$ depends on a fixed parameter $\bdr^{(n)}$ obtained in previous iteration, we denote it as $G_{R}^{(n)}(\lambda)$ for convenience. 
We consider the case where $\lambda^{(n)}$ is not the exact solution of the monotonic function $G_{R}^{(n)}(\lambda)$ due to numerical inaccuracy, i.e., the Newton's method for the root finding subroutine of $G_{R}^{(n)}(\lambda)=0$ stops with some inaccuracy, i.e., $|G_{R}^{(n)}(\lambda^{(n)})|\leq \delta$, where $\delta$ is the tolerance.

Then, similar to the derivation of Lemma \ref{lem: 1}, we have the estimation
\begin{multline*}
f_{R}(\bdw^{(n)}, \bdr^{(n)})-f_{R}(\bdw^{*}, \bdr^{(n)})=f_{R}(\bdw^{(n)}, \bdr^{(n)})+\lambda^{(n)} D-f_{R}(\bdw^{*}, \bdr^{(n)})-\lambda^{(n)} D \\
=\left(f_{R}(\bdw^{(n)}, \bdr^{(n)})\!+\!\lambda^{(n)}\sum_{i,j}p_{i}w_{ij}^{(n)}d_{ij}\!-\!\lambda^{(n)} G_{R}^{(n)}(\lambda^{(n)})\right)\!
\!-\!\!\left(f_{R}(\bdw^{*}, \bdr^{(n)})\!+\!\lambda^{(n)}\sum_{i,j}p_{i}w_{ij}^{*}d_{ij}\right) \\
=\sum_{i,j}p_{i}w_{ij}^{(n)}\left[\log \frac{e^{-\lambda^{(n)}d_{ij}}}{\sum_{k} r_{k}^{(n)}e^{-\lambda^{(n)} d_{ik}}}\!+\!\lambda^{(n)}d_{ij} \right]\!-\!\sum_{i,j} p_{i}w_{ij}^{*}\log\frac{w_{ij}^{*}}{r_{j}^{(n)}e^{-\lambda^{(n)}d_{ij}}}\!-\!\lambda^{(n)} G_{R}^{(n)}(\lambda^{(n)}) \\
=\sum_{i}\left[\sum_{j}p_{i}w_{ij}^{(n)}\right]\log\frac{1}{\sum_{k} r_{k}^{(n)}e^{-\lambda^{(n)} d_{ik}}}-\sum_{i,j} p_{i}w_{ij}^{*}\log\frac{w_{ij}^{*}}{r_{j}^{(n)}e^{-\lambda^{(n)}d_{ij}}}-\lambda^{(n)} G_{R}^{(n)}(\lambda^{(n)}) \\
=\sum_{i}\left[\sum_{j}p_{i}w_{ij}^{*}\right]\log\frac{1}{\sum_{k} r_{k}^{(n)}e^{-\lambda^{(n)} d_{ik}}}-\sum_{i,j} p_{i}w_{ij}^{*}\log\frac{w_{ij}^{*}}{r_{j}^{(n)}e^{-\lambda^{(n)}d_{ij}}}-\lambda^{(n)} G_{R}^{(n)}(\lambda^{(n)}) \\
=-\sum_{i=1}^{M} p_{i} D_{KL}\left(\bdw_{i}^{*}\big\|\bdw_{i}^{(n)}\right)-\lambda^{(n)} G_{R}^{(n)}(\lambda^{(n)}) \leq -D_{KL}\left(\sum_{i=1}^{M}p_{i}\bdw_{i}^{*}\Big\|\sum_{i=1}^{M}p_{i}\bdw_{i}^{(n)}\right)+\lambda^{(n)}\delta.
\end{multline*}
Thus, using the result obtained in Lemma \ref{lem: 2}, we have the estimation
\begin{multline*}
f_{R}(\bdw^{(n)},\bdr^{(n)})-f_{R}(\bdw^{*},\bdr^{*})=\left(f_{R}(\bdw^{(n)},\bdr^{(n)})-f_{R}(\bdw^{*},\bdr^{(n)})\right)+\left(f_{R}(\bdw^{*},\bdr^{(n)})-f_{R}(\bdw^{*},\bdr^{*})\right) \\
\leq-D_{KL}\Big(\sum_{i}p_{i}\bdw_{i}^{*}\Big\|\sum_{i}p_{i}\bdw_{i}^{(n)}\Big)+ D_{KL}\Big(\bdr^{*}\big\|\bdr^{(n)}\Big)+\lambda^{(n)}\delta \\
= -D_{KL}\left(\bdr^{*}\Big\|\bdr^{(n+1)}\right)+D_{KL}\left(\bdr^{*}\Big\|\bdr^{(n)}\right)+\lambda^{(n)}\delta.
\end{multline*}
Then, by summing up these inequalities for all indexes, we have
\begin{equation*}
\sum_{k=1}^{n}\Big(f_{R}(\bdw^{(k)},\bdr^{(k)})-f_{R}(\bdw^{*},\bdr^{*})\Big)\leq D_{KL}(\bdr^{*}\|\bdr^{(1)})-D_{KL}(\bdr^{*}\|\bdr^{(n+1)})+\big(\sum_{k=1}^{n}\lambda^{(k)}\big)\delta.
\end{equation*}
Hence, we have the estimation 
\begin{multline*}
f_{R}(\bdw^{(n)},\bdr^{(n)})-f_{R}(\bdw^{*},\bdr^{*}) \leq \frac{1}{n}\left[ D_{KL}(\bdr^{*}\|\bdr^{(1)})-D_{KL}(\bdr^{*}\|\bdr^{(n+1)})+\big(\sum_{k=1}^{n}\lambda^{(k)}\big)\delta \right] \\
\leq \frac{1}{n}\left[ D_{KL}(\bdr^{*}\|\bdr^{(1)})+\big(\sum_{k=1}^{n}\lambda^{(k)}\big)\delta \right] \leq \frac{1}{n} \left[
\log N + \big(\sum_{k=1}^{n}\lambda^{(k)}\big)\delta \right].
\end{multline*}
Here, we use the fact that $r^{(1)}_j=1/N$ and $\sum_j r_j^*\log r_j^*\leq 0$. 

Then, we show that the sequence $\lambda^{(k)}$, $k=1,2\cdots$ are bounded by some constant $C$.
Apparently, $|G_{R}^{(n)}(\lambda^{(n)})|\leq \delta$ implies 
\begin{equation*}
    \sum_{i=1}^{M}\sum_{j=1}^{N} d_{ij} p_{i} \left(e^{-\lambda^{(n)} d_{ij}}r_{j}^{(n)}\bigg/\left(\sum_{k=1}^{N} e^{-\lambda^{(n)} d_{ik}}r_{k}^{(n)}\right)\right)\geq D-\delta.
\end{equation*}
We assume that $\lambda^{(n)}$ is not bounded, and then there exists a subsequence $\lambda^{(n_k)}\rightarrow +\infty$. 
For vector $\bm{r}^{(n_k)}$, they are bounded sequence since $\|\bm{r}^{(n_k)}\|_1=1$. 
Therefore, there exists a convergent subsequence $\bm{r}^{({n_k}_t)}\rightarrow \bm{r}^{*}$. 
For simplicity, we rewrite the index ${n_k}_t$ as $n$, and then 
\begin{equation*}
    \begin{aligned}
         D-\delta&\leq \sum_{i=1}^{M}\sum_{j=1}^{N} d_{ij} p_{i} \left(e^{-\lambda^{(n)} d_{ij}}r_{j}^{(n)}\bigg/\left(\sum_{k=1}^{N} e^{-\lambda^{(n)} d_{ik}}r_{k}^{(n)}\right)\right)\\
          & =\sum_{i=1}^{M}p_{i}\left(\sum_{j=1}^{N} d_{ij}  e^{-\lambda^{(n)} d_{ij}}r_{j}^{(n)}\right)\bigg/\left(\sum_{k=1}^{N} e^{-\lambda^{(n)} d_{ik}}r_{k}^{(n)}\right)\\
          &=\sum_{i=1}^{M}p_{i}\left(\sum_{j=1}^{N} d_{ij}  e^{-\lambda^{(n)} (d_{ij}-\min_t d_{it})}r_{j}^{(n)}\right)\bigg/\left(\sum_{k=1}^{N} e^{-\lambda^{(n)} (d_{ik}-\min_t d_{it})}r_{k}^{(n)}\right).
    \end{aligned}
\end{equation*}
If $j$ satisfies $d_{ij}=\min_t d_{it}$, then $e^{-\lambda^{(n)} (d_{ij}-\min_t d_{it})}r_{j}^{(n)}\rightarrow r_j^*$, as $n \rightarrow +\infty$. 
If $j$ satisfies $d_{ij}>\min_t d_{it}$, then $e^{-\lambda^{(n)} (d_{ij}-\min_t d_{it})}r_{j}^{(n)}\rightarrow 0$, as $n \rightarrow +\infty$. 
So by taking the limit $n \rightarrow +\infty$, we have 
\begin{equation*}
    D-\delta\leq \sum_{i=1}^{M}p_{i}\left(\sum_{j:d_{ij}=\min_t d_{it}} d_{ij} r_{j}^{(n)}\right)\bigg/\left(\sum_{j:d_{ij}=\min_t d_{it}} r_{k}^{(n)}\right),
\end{equation*}
where $\sum_{j:d_{ij}=\min_t d_{it}}$ means taking the sum of terms whose index $j$ satisfies $d_{ij}=\min_t d_{it}$. 
Thus, we get $D-\delta\leq \sum_{i=1}^M p_i \min_t d_{it}=D_{\min}$, which is contradictory to $D\in (D_{\min},D_{\max})$, since $\delta$ is sufficiently small. 
Hence, we have proved that the sequence $\lambda^{(k)},k=1,2\cdots$ are bounded by some constant $C$. 
Consequently, to ensure $\varepsilon$-approximation, it is sufficient for the iteration number $n$ and the stopping tolerance $\delta$ to satisfy:  
\begin{equation}                                            \frac{\log N}{n} \leq \frac{\varepsilon}{2}, \quad\quad \frac{(\sum_{k=1}^{n}\lambda^{(k)})}{n} \delta \leq C\delta \leq \frac{\varepsilon}{2}.  
\end{equation}
Here, the constant $C$ is the upper bound of $\lambda^{(k)}$. 
Hence, to ensure $\varepsilon$-approximation, the iteration number of the proposed CBA algorithm satisfies $n\geq \frac{2\log N}{\varepsilon}$.
Note that the multiplications in the CBA algorithm and the Newton's method require $O(MN)$ arithmetic operations in each iteration step, and the Newton's method achieves $\delta$ accuracy in $\log|\log \delta|$ steps due to the locally second order convergence property \cite{nesterov2018lectures}.
Thus, the total arithmetic operations of the CBA algorithm can be upper bounded by 
\begin{equation*}
    O\left(\frac{2\log N}{\varepsilon} \left(O(MN)+O(MN) \log|\log\frac{\varepsilon}{C}|\right)\right).
\end{equation*}
Hence, we prove the total arithmetic operations of $O\left(\frac{MN\log N}{\varepsilon}(1+\log|\log \varepsilon|)\right)$.
\end{proofs}

\subsection{Convergence analysis for the DR function} \label{subsec_3_DR}
First, we present the proof of Theorem \ref{thm: 2}.
Since the monotonic function $G_{D}(\lambda)$ depends on the parameter $\bdr^{(n)}$ from the previous iteration, we denote it as $G_{D}^{(n)}(\lambda)$ for convenience.

\begin{proofs}
Given $\bdr^{(n)}$, denote $\lambda^{(n)}$ as the root of the monotonic function $G_{D}^{(n)}(\lambda)$, where
\begin{equation*}
    G_{D}^{(n)}(\lambda) \triangleq -\sum_{i=1}^M p_i \log\left(\sum_{j=1}^N r_j^{(n)} e^{-\lambda d_{ij}}\right) -\lambda\sum_{i=1}^{M}\sum_{j=1}^{N} d_{ij} p_{i} \left(e^{-\lambda d_{ij}}r_{j}^{(n)}\bigg/\left(\sum_{j=1}^{N} e^{-\lambda d_{ij}}r_{j}^{(n)}\right)\right)-R.
\end{equation*}
Then, we have the estimation after the $n_{th}$ iteration, 
\begin{multline*}
f_{D}(\bdw^{(n)})-f_{D}(\bdw^*)=f_{D}(\bdw^{(n)})+R/\lambda^{(n)}-f_{D}(\bdw^*)- R/\lambda^{(n)} \\
=\left(f_{D}(\bdw^{(n)})\!+\!\frac{1}{\lambda^{(n)}}\sum_{i,j}p_{i}w_{ij}^{(n)}\log\frac{w_{ij}^{(n)}}{r_j^{(n)}}\right)\!\!-\left(f_{D}(\bdw^*)\!+\!\!\frac{1}{\lambda^{(n)}}\sum_{i,j}p_{i}w_{ij}^{*}\log\frac{w_{ij}^{*}}{r_j^{*}}\right) \\
=\frac{1}{\lambda^{(n)}} \left[\sum_{i,j}p_{i}w_{ij}^{(n)}\left[\log\left(\frac{r_{j}^{(n)}e^{-\lambda^{(n)}d_{ij}}}{\sum_{k}r_{k}^{(n)}e^{-\lambda^{(n)}d_{ik}}}\right)-\log\left({r_{j}^{(n)}e^{-\lambda^{(n)} d_{ij}}}\right)\right]-\sum_{i,j}p_{i}w_{ij}^{*}\log\frac{w_{ij}^{*}}{r_{j}^{*}e^{-\lambda^{(n)} d_{ij}}} \right] \\
=\frac{1}{\lambda^{(n)}} \left[\sum_{i}p_{i}\left(\sum_j w_{ij}^{(n)}\right)\log\frac{1}{\sum_j r_{j}^{(n)}e^{-\lambda^{(n)} d_{ij}}}-\sum_{i,j}p_{i}w_{ij}^{*}\log\frac{w_{ij}^{*}}{r_{j}^{(n)}e^{-\lambda^{(n)} d_{ij}}}+\sum_{i,j}p_i w_{ij}^* \log\frac{r_j^{*}}{r_j^{(n)}} \right] \\
=\frac{1}{\lambda^{(n)}}\left[\sum_{i}p_{i}\left(\sum_j w_{ij}^{*}\right)\log\frac{1}{\sum_j r_{j}^{(n)}e^{-\lambda^{(n)} d_{ij}}}-\sum_{i,j}p_{i}w_{ij}^{*}\log\frac{w_{ij}^{*}}{r_{j}^{(n)} e^{-\lambda^{(n)} d_{ij}}}+\sum_{j}r_j^{*} \log \frac{r_j^{*}}{r_j^{(n)}} \right] \\
=\frac{1}{\lambda^{(n)}}\left[-\sum_{i,j}p_{i}w_{ij}^{*}\log\frac{w_{ij}^{*}\sum_j r_{j}^{(n)}e^{-\lambda^{(n)} d_{ij}}}{r_{j}^{(n)}e^{-\lambda^{(n)} d_{ij}}}+\sum_{j}r_j^{*} \log \frac{r_j^{*}}{r_j^{(n)}}\right] \\
=\frac{1}{\lambda^{(n)}}\left[-\sum_{i,j}p_{i}w_{ij}^{*}\log\frac{w_{ij}^{*}}{w_{ij}^{(n)}}+\sum_{j}r_j^* \log\frac{r_j^{*}}{r_j^{(n)}}\right] \\
\leq \frac{1}{\lambda^{(n)}}\left[-D_{KL}\Big(\sum_{i}p_{i}\bdw^{*}_i\| \sum_{i}p_{i}\bdw^{(n)}_i\Big) + D_{KL}(\bdr^*\|\bdr^{(n)})\right] \\
\leq C \Big(D_{KL}(\bdr^*\|\bdr^{(n)})-D_{KL}(\bdr^*\|\bdr^{(n+1)})\Big).
\end{multline*}
The first inequality holds in the above derivation, due to the convex property of the KL divergence.
The constant $C$ denotes a uniform upper bound of the sequence $\{\frac{1}{\lambda^{(n)}}\}$, due to the strict positivity of the sequence $\{\lambda^{(n)}\}$. 
In fact, we can prove that the sequence $\{\lambda^{(n)}\}$ is lower bounded, since
\begin{multline*}
    0=G_{D}^{(n)}(\lambda)=-\sum_{i=1}^M p_i \log\left(\sum_{j=1}^N r_j^{(n)} e^{-\lambda d_{ij}}\right) -\lambda\sum_{i=1}^{M}\sum_{j=1}^{N} d_{ij} p_{i} \frac{e^{-\lambda d_{ij}}r_{j}^{(n)}}{\sum_{j=1}^{N} e^{-\lambda d_{ij}}r_{j}^{(n)}}-R \\
    \leq -\sum_{i=1}^M p_i \log\left(\sum_{j=1}^N r_j^{(n)} e^{-\lambda d_{ij}}\right)-R \leq -\sum_{i=1}^M p_i \log\left(\sum_{j=1}^N r_j^{(n)} e^{-\lambda \max_j d_{ij}}\right)-R\\
    =-\sum_i p_i \log e^{-\lambda \max_j d_{ij}}-R=\lambda\sum_i p_i \max_j d_{ij}-R.
\end{multline*}
The above inequality means that $\lambda\geq R/\left(\sum_i p_i \max_j d_{ij}\right)>0$. 
Hence, the sequence $\{\lambda^{(n)}\}$ is lower bounded.
Then, by summing up these inequalities for $\bdw^{(1)},\cdots, \bdw^{(n)}$, we have the estimation
\begin{multline*}
\sum_{k=1}^{n}\Big(f_{D}(\bdw^{(k)})-f_{D}(\bdw^{*})\Big)\leq C \left(D_{KL}(\bdr^{*}\|\bdr^{(1)})-D_{KL}(\bdr^{*}\|\bdr^{(n+1)}) \right) \\
\leq C \left( D_{KL}(\bdr^{*}\|\bdr^{(1)})\right) := C_{const},
\end{multline*}
where $C_{const}$ is a constant estimated as $O(\log N)$, when taking uniform initialization.
Next, we prove the non-increasing property of $f_{D}(\bdw)$ during iterations. In the above analysis, we substitute $\bdw^*$ and $\bdr^*$ by $\bdw^{(n-1)}$ and $\bdr^{(n-1)}$, and note that $\sum_i p_i w_{ij}^{(n-1)}=r_j^{(n)} $. Similarly, we have 
\begin{equation*}
    f_{D}(\bdw^{(n)})-f_{D}(\bdw^{(n-1)})=-\frac{1}{\lambda^{(n)}}\sum_{i}p_{i}D_{KL}(\bdw^{(n-1)}_i\| \bdw^{(n)}_i)-\frac{1}{\lambda^{(n)}} D_{KL}(\bdr^{(n)}\|\bdr^{(n-1)})\leq 0.
\end{equation*}
Since $f_{D}(\bdw^{(k)})\leq f_{D}(\bdw^{(k-1)})$, we obtain
\begin{equation*}
    0\leq n\left( f_{D}(\bdw^{(n)})-f_{D}(\bdw^{*})\right)\leq \sum_{k=1}^{n}\Big(f_{D}(\bdw^{(k)})-f_{D}(\bdw^{*})\Big)\leq C_{const},
\end{equation*}
i.e., $f_{D}(\bdw^{(n)})-f_{D}(\bdw^{*})\simeq O(1/n)$. 
\end{proofs}

\vspace{+.1in}
Furthermore, we provide a computational complexity analysis of the CBA algorithm for approximating the optimal solution in the DR function with accuracy $\varepsilon$. 
In this regard, we present the proof of Theorem \ref{thm: 4}. 
\begin{proofs}
Similar to the previous proof, we use the notation $G_{D}^{(n)}(\lambda)$ for convenience. 
Here, we consider the case where $\lambda^{(n)}$ is not the exact solution of the monotonic function $G_{D}^{(n)}(\lambda)$ due to numerical inaccuracy.
In other words, the Newton's method for the root finding subroutine of $G_{D}^{(n)}(\lambda)=0$ stops with $|G_{D}^{(n)}(\lambda^{(n)})|\leq \delta$, where $\delta$ is the tolerance.

Then, similar to the derivation process in Theorem \ref{thm: 2}, we have the estimation
\begin{multline*}
f_{D}(\bdw^{(n)})-f_{D}(\bdw^{*})=f_{D}(\bdw^{(n)})+\frac{1}{\lambda^{(n)}} R-f_{D}(\bdw^{*})-\frac{1}{\lambda^{(n)}} R \\
=\frac{1}{\lambda^{(n)}}\left[\left(\lambda^{(n)}\sum_{i,j}p_{i}w_{ij}^{(n)}d_{ij}\!+\!\sum_{i,j}p_{i}w_{ij}^{(n)}\log\frac{w_{ij}^{(n)}}{r_j^{(n)}}\!-\!G_{D}^{(n)}(\lambda^{(n)})\right)\!\!\!-\!\!\!\left(\lambda^{(n)}\sum_{i,j}p_{i}w_{ij}^{*}d_{ij}\!+\!\!\sum_{i,j}p_{i}w_{ij}^{*}\log\frac{w_{ij}^{*}}{r_j^{*}}\right) \right]\\
=\frac{1}{\lambda^{(n)}}\left[\sum_{i}\left(\sum_{j}p_{i}w_{ij}^{(n)}\right)\log\frac{1}{\sum_{k} r_{k}^{(n)}e^{-\lambda^{(n)} d_{ik}}}-\sum_{i,j} p_{i}w_{ij}^{*}\log\frac{w_{ij}^{*}}{r_{j}^{(n)}e^{-\lambda^{(n)}d_{ij}}} - G_{D}^{(n)}(\lambda^{(n)}) + \sum_j r_j^*\log \frac{r_j^*}{r_j^{(n)}} \right] \\
=\frac{1}{\lambda^{(n)}}\left[\sum_{i}\left(\sum_{j}p_{i}w_{ij}^{*}\right)\log\frac{1}{\sum_{k} r_{k}^{(n)}e^{-\lambda^{(n)} d_{ik}}}\!-\!\sum_{i,j} p_{i}w_{ij}^{*}\log\frac{w_{ij}^{*}}{r_{j}^{(n)}e^{-\lambda^{(n)}d_{ij}}}\!-\!G_{D}^{(n)}(\lambda^{(n)})\!+\!D_{KL}(\bm{r}^*\|\bm{r}^{(n)})\right] \\
=-\frac{1}{\lambda^{(n)}}\sum_{i,j} p_{i}w_{ij}^{*}\log\frac{w_{ij}^{*}{\sum_{k} r_{k}^{(n)}e^{-\lambda^{(n)} d_{ik}}}}{r_{j}^{(n)}e^{-\lambda^{(n)}d_{ij}}}-\frac{1}{\lambda^{(n)}} G_{D}^{(n)}(\lambda^{(n)}) +\frac{1}{\lambda^{(n)}} D_{KL}(\bm{r}^*\|\bm{r}^{(n)})\\
=-\frac{1}{\lambda^{(n)}}\sum_{i=1}^{M} p_{i} D_{KL}\left(\bdw_{i}^{*}\big\|\bdw_{i}^{(n)}\right)-\frac{1}{\lambda^{(n)}} G_{D}^{(n)}(\lambda^{(n)})+\frac{1}{\lambda^{(n)}} D_{KL}(\bm{r}^*\|\bm{r}^{(n)})\\
\leq -\frac{1}{\lambda^{(n)}}D_{KL}\left(\sum_{i=1}^{M}p_{i}\bdw_{i}^{*}\Big\|\sum_{i=1}^{M}p_{i}\bdw_{i}^{(n)}\right)+\frac{1}{\lambda^{(n)}} D_{KL}(\bm{r}^*\|\bm{r}^{(n)})+\frac{1}{\lambda^{(n)}} \delta.
\end{multline*}
Thus, we have the estimation
\begin{multline*}
f_{D}(\bdw^{(n)})-f_{D}(\bdw^{*})
\leq \frac{1}{\lambda^{(n)}}\left(-D_{KL}(\bm{r}^*\Big\|\bm{r}^{(n+1)})+D_{KL}(\bm{r}^*\|\bm{r}^{(n)})\right)+\frac{1}{\lambda^{(n)}} \delta\\
\leq C\left(-D_{KL}\left(\bm{r}^*\Big\|\bm{r}^{(n+1)}\right)+D_{KL}(\bm{r}^*\|\bm{r}^{(n)})\right)+C \delta.
\end{multline*}
Here, we also denote by $C$ a constant upper bound for $1/\lambda^n$, similar to the analysis in the proof of Theorem \ref{thm: 2}. 
Then, by summing up these inequalities for $n=1,2\cdots$, we have
\begin{equation*}
\sum_{k=1}^{n}\Big(f_{D}(\bdw^{(k)})-f_{D}(\bdw^{*})\Big)\leq C\Big(D_{KL}(\bdr^{*}\|\bdr^{(1)})-D_{KL}(\bdr^{*}\|\bdr^{(n+1)})\Big)+nC\delta.
\end{equation*}
Hence, we have the estimation 
\begin{multline*}
f_{D}(\bdw^{(n)})-f_{D}(\bdw^{*})\leq \frac{1}{n}\left[ C\cdot D_{KL}(\bdr^{*}\|\bdr^{(1)})-C\cdot D_{KL}(\bdr^{*}\|\bdr^{(n+1)})+nC\delta \right] \\
\leq \frac{1}{n}\left[ C\cdot D_{KL}(\bdr^{*}\|\bdr^{(1)})+nC\delta \right] \leq \frac{1}{n} \left[
C\log N + nC\delta \right].
\end{multline*}
Here, we use the fact that $r^{(1)}_j=1/N$ and $\sum_j r_j^*\log r_j^*\leq 0$.
Hence, to ensure $\varepsilon$-approximation, it suffices to take the condition of iteration number $n$ and the stopping tolerance $\delta$ as 
\begin{equation}
    C\frac{\log N}{n} \leq \frac{\varepsilon}{2}, \quad C \delta\leq \frac{\varepsilon}{2}.  
\end{equation}
Note that the Newton's method only needs $O(MN)$ arithmetic operations in each iteration step and it obtains $\delta$-accuracy in $\log|\log \delta|$ steps due to the locally second order convergence property \cite{nesterov2018lectures}, hence similar to the analysis in the proof of Theorem \ref{thm: 3}, we have the arithmetic operations of $O\left(\frac{MN\log N}{\varepsilon}(1+\log|\log \varepsilon|)\right)$.
\end{proofs}

\section{Numerical Results and Discussions} \label{Sec_4_numerical}
This section evaluates the performance of the CBA algorithm by conducting some numerical experiments. 
These experiments have been implemented by Matlab R2022a on a Linux platform with 128G RAM and one Intel(R) Xeon(R) Gold 5117 CPU@2.00GHz. 

\subsection{Accuracy and Efficiency}
To illustrate the accuracy and efficiency of the CBA algorithm, we establish a comparative framework by benchmarking its performance against the original BA algorithm \cite{blahut1972computation} as a baseline. 
Additionally, we extend our comparison to include two recently developed algorithms: the Alternating Sinkhorn (AS) algorithm \cite{wu2022communication}, which relies on an entropy regularized Optimal Transport (OT) framework for the RD function, and a Bregman divergence based EM algorithm \cite{hayashi2023bregman}.

We conduct our study on two specific sources: the Gaussian source with squared error distortion \cite{book_element}, and the Laplacian source with absolute error distortion\cite{berger1971}. 
These sources fortunately admit closed-form expressions for their RD functions, and are chosen here to illustrate and validate the accuracy and efficiency of the CBA algorithm when compared to other existing methods. 
Since all the studied algorithms apply to discrete sources, we consider discretized versions of these sources when conducting our numerical experiments.
The Gaussian sources, for instance, are frequently encountered 
in scenarios involving Discrete Fourier Transform (DFT), making it a fundamental component in signal processing and spectrum analysis \cite{pearlman1978source}.
Laplacian sources find applications in fields like speech signal compression \cite{eltoft2006multivariate,sitaula2022neonatal}, serving as effective models for the statistical characteristics of speech signals, and in neural network weight compression, a field gaining prominence in the realm of machine learning and deep learning \cite{han2015deep,peric2021robust}.
Hence, our study focuses on the discretized Gaussian and Laplacian sources using the CBA algorithm due to their great importance, aiming to demonstrate its accuracy and efficiency. 
Notably only the CBA and the BA algorithms possess the capability to concurrently compute both the RD and DR functions, while there has been no prior research on utilizing the AS and the EM algorithms for the resolution of the DR functions.
In order to apply these algorithms, we truncate the continuous probability distribution of the input source $X$ within an interval $[-L,L]$ and discretize the interval using a set of uniform grid points $\{x_{i}\}_{i=1}^{K}$:
\begin{equation*}
x_{i}=-L+(i-1/2) \delta,\quad \delta=2{L}/{K},~i=1,\cdots,K.
\end{equation*}
Similarly, we can discretize the the continuous probability distribution of the output source $Y$ within an interval $[-L,L]$ in the following way:
\begin{equation*}
y_{j}=-L+(j-1/2) \tilde{\delta},\quad \tilde{\delta}=2{L}/{N},~j=1,\cdots,N.
\end{equation*}
We set $L = 8$, $K = 100$ and $N=K$ for these sources.
Moreover, we let the Gaussian source have mean zero and variance $\sigma = 1$, and the Laplacian source have mean zero and scale parameter $b = 1$.
The computational results are summarized in Table \ref{table_RD} (RD) and Table \ref{table_DR} (DR), wherein each result is obtained by repeating the experiment $100$ times. 
\begin{table}[ht] 
	\renewcommand\arraystretch{1.1}
	\centering
	\caption{Comparison between CBA, AS, EM and BA algorithms for the RD function.} 
    \vspace{+.1in}
	\setlength{\tabcolsep}{.6 mm}{
        \scalebox{1.0}{
		\begin{tabular}{c|c|c|c|c}
			\toprule
			\multirow{2}{*}{($D$,\ $R$,\ Slope $\lambda$)} & \multicolumn{4}{c}{(Time in sec, Number of iterations)} \\
			\cline{2-5}
		      & CBA & BA & AS & EM \\ 
		  \hline 
            \multicolumn{5}{c}{Discretized Gaussian Source} \\
            \hline 
			$(0.1,\ 1.1513,\ 5.0000)$ & $(0.0071,\ 8)$ & $(0.2158,\ 14)$ & $(0.0891,\ 182)$ & $(0.0336,\ 8)$  \\
		      $(0.3,\ 0.6020,\ 1.6667)$ & $(0.0111,\ 16)$ & $(0.4101,\ 32)$ & $(0.1610,\ 388)$ & $(0.0634,\ 16)$  \\
                $(0.5,\ 0.3466,\ 1.0000)$ & $(0.0178,\ 27)$ & $(0.8098,\ 69)$ & $(0.2637,\ 657)$ & $(0.1061,\ 27)$  \\
                $(0.7,\ 0.1783,\ 0.7143)$ & $(0.0340,\ 52)$ & $(2.9356,\ 191)$ & $(0.2376,\ 538)$ & $(0.1951,\ 52)$  \\
                $(0.9,\ 0.0527,\ 0.5556)$ & $(0.1023,\ 164)$ & $(18.9542,\ 1445)$ & $(0.1353,\ 258)$ & $(0.6290,\ 164)$  \\
            \hline 
            \multicolumn{5}{c}{Discretized Laplacian Source} \\
            \hline 
			$(0.1,\ 2.1530,\ 7.8059)$ & $(0.0197,\ 43)$ & $(0.9561,\ 89)$ & $(1.1421,\ 2539)$ & $(0.1742,\ 43)$  \\
		      $(0.3,\ 1.1797,\ 3.1924)$ & $(0.2609,\ 649)$ & $(13.6523,\ 1420)$ & $(0.2180,\ 453)$ & $(2.4794,\ 649)$  \\
                $(0.5,\ 0.6830,\ 1.9671)$ & $(1.1867,\ 2783)$ & $(67.3326,\ 6515)$ & $(1.1313,\ 2741)$ & $(11.1018,\ 2781)$  \\
                $(0.7,\ 0.3506,\ 1.4161)$ & $(2.7466,\ 6493)$ & $(141.6602,\ 13743)$ & $(2.3278,\ 6465)$ & $(25.4069,\ 6490)$  \\
                $(0.9,\ 0.1010,\ 1.1047)$ & $(4.9143,\ 11437)$ & $(225.9824,\ 22163)$ & $(4.0589,\ 11412)$ & $(46.7251,\ 11429)$  \\
            \bottomrule
	\end{tabular}}

\vspace{+.1in}

\footnotesize{{Notes: a) Columns 2-5 are the average computing time (in seconds) and the number of iterations. b) The BA algorithm cannot compute the RD function directly with a given target $D$, and hence we adaptively search for the slope $\lambda$ to ensure accuracy. It generally takes about $50$ trials to find the corresponding $\lambda$. In the column for the BA algorithm, the time is measured including all the trials, whereas the number of iterations is only for the last trial with the found $\lambda$. c) The stopping criterion is defined as the reduction from the previous step being less than $10^{-10}$. 
}}}
\label{table_RD}
\end{table}
From the outcomes presented in Table \ref{table_RD}, the CBA algorithm excels in computational time compared to the original BA algorithm for the investigated two sources. Notably, the CBA algorithm frequently achieves speed-up ratios exceeding $30$. 
Moreover, in comparison to the EM algorithm, the CBA algorithm significantly accelerates computational speed, particularly evident for the discretized Laplacian source. In this case, the CBA algorithm achieves speed-up ratios of approximately $10$.
Although the computational time of the CBA and AS algorithms are sometimes comparable in some scenarios, it is crucial to emphasize that the CBA algorithm offers a theoretical convergence guarantee, as proved by Theorem \ref{thm: 1}, while the AS algorithm not. 

Furthermore, as illustrated in Table \ref{table_DR}, the CBA algorithm demonstrates superior computational efficiency in comparison to the BA algorithm for the two sources under investigation. 
In our observed data, the CBA algorithm consistently demonstrates a reduced iteration count and frequently achieves significant speed-up ratios exceeding $40$.
It should be noted that the utilization of the AS and EM algorithms for direct computation of the DR function remains currently unknown, hence the computational time results for these approaches are omitted from the table.

\begin{table}[ht] 
	\renewcommand\arraystretch{1.1} 
	\centering
	\caption{Comparison between CBA and BA algorithms for the DR function.} 
    \vspace{+.1in}
	\setlength{\tabcolsep}{1.3 mm}{
        \scalebox{1.0}{
		\begin{tabular}{c|c|c|c}
		\toprule
            \multirow{2}{*}{Source} & \multirow{2}{*}{($D$,\ $R$,\ Slope $\lambda$)} & \multicolumn{2}{c}{(Time in sec, Number of iterations)} \\
			\cline{3-4}
		      &  & CBA & BA \\ 
			\hline 
            \multirow{5}{*}{Discretized  Gaussian} 
			& $(0.8187,\ 0.1,\ 0.6107)$ & $(0.0583,\ 96)$ & $(4.6985,\ 505)$ \\
			& $(0.5488,\ 0.3,\ 0.9111)$ & $(0.0213,\ 34)$ & $(0.7325,\ 86)$ \\
                & $(0.3679,\ 0.5,\ 1.3591)$ & $(0.0136,\ 20)$ & $(0.3842,\ 41)$ \\
                & $(0.2466,\ 0.7,\ 2.0276)$ & $(0.0104,\ 15)$ & $(0.2372,\ 25)$ \\
                & $(0.1653,\ 0.9,\ 3.0248)$ & $(0.0085,\ 11)$ & $(0.1852,\ 18)$ \\
            \hline 
			\multirow{5}{*}{Discretized Laplacian} 
			& $(0.9009,\ 0.1,\ 1.1036)$ & $(4.4983,\ 11085)$ & $(159.9343,\ 19735)$ \\
			& $(0.6019,\ 0.5,\ 1.6421)$ & $(1.5678,\ 3915)$ & $(68.0310,\ 8466)$ \\
                & $(0.4006,\ 0.9,\ 2.4338)$ & $(0.5072,\ 1243)$ & $(23.5057,\ 2758)$ \\
                & $(0.2644,\ 1.3,\ 3.5822)$ & $(0.1526,\ 396)$ & $(6.0424,\ 802)$ \\
                & $(0.1714,\ 1.7,\ 5.2095)$ & $(0.0510,\ 116)$ & $(2.1416,\ 254)$ \\
			\bottomrule
	\end{tabular}}
 
\vspace{+.1in}

\footnotesize{{Notes: a) The settings are the same as those in Table \ref{table_RD}.}}} 
\label{table_DR} 
\end{table}

\subsection{Convergence Behaviour Near Bifurcations}

Here, we consider a case with bifurcation structure and linear segment in the RD function.
The source distribution and its associated distortion measure are given by \cite[Example 2.7.3]{berger1971}\cite{agmon2022root}
\begin{equation*}
P_{X}=(0.4,0.6) \text{ and } d(x, \hat{x})=\left(\begin{array}{ccc}
1 & 0 & 0.3 \\
0 & 1 & 0.3
\end{array}\right).
\end{equation*}
In this case, the curve of the RD function (shown in the left sub-graph of Figure 2) exhibits two bifurcations at $D_{1} \approx 0.14$ and $D_{2} \approx 0.26$, which leads to a significant negative impact on the convergence of computation algorithms, especially for target distortions close to these two bifurcations.   
Moreover, the RD curve contains a linear segment between the two bifurcations, over which the slope does not change, and thus the BA algorithm cannot produce the entire linear segment since for each given slope it ends up with a single point on the RD curve. As shown in the left sub-graph of Figure 2, the CBA algorithm successfully produces the entire RD curve without accuracy loss.

\begin{figure}[ht]
    \centerline{\includegraphics[width=0.95\textwidth]{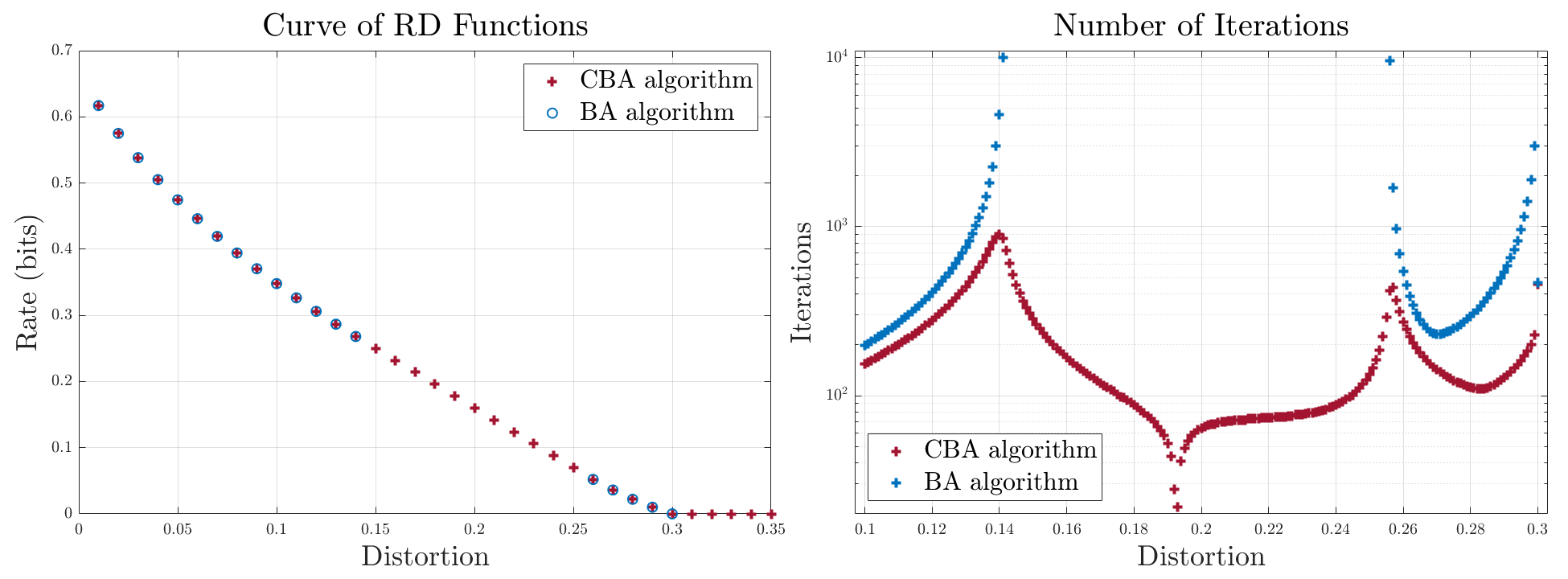}}
    \label{Fig: baff}
   \caption{The RD curve produced by the CBA algorithm and the BA algorithm (Left), and comparison between the numbers of iterations (Right). Here, the stop condition is set as that the decrease from the previous step is below $10^{-8}$. For each given target distortion $D$, the BA algorithm is executed using the slope $\lambda$ already computed by the CBA algorithm with high accuracy as the predetermined parameter.}
\end{figure}

To further highlight the advantage of the CBA algorithm over the BA algorithm, we compare the number of iterations required for convergence with varying target distortions, displayed in the right sub-graph of Figure 2.
The results demonstrate that the CBA algorithm requires fewer iterations than the BA algorithm. Particularly, the BA algorithm cannot sweep out the linear segment of the RD curve between the two bifurcations, but the CBA algorithm still performs well there.
Furthermore, even near the bifurcations, where the support of the reproduction probability distribution abruptly changes, the CBA algorithm still achieves convergence within 1000 iterations, saving more than $90\%$ of those required by the BA algorithm. 
By updating the multiplier $\lambda$ in each iteration, the CBA algorithm avoids problems of singularity near bifurcations, leading to a much smaller number of iterations.

\subsection{Phase Transitions in the Uniform Source}

To examine the efficiency of the CBA algorithm in scenarios favoring discrete optimal reproduction, we concentrate on instances characterized by squared error distortion measures. 
Specifically, our focus gravitates towards the uniform source, known for its discrete optimal reproduction characteristics (see, e.g., \cite{rose94}). 
As demonstrated in \cite{rose94}, under squared error distortion, the optimal reproduction probability distribution is typically purely discrete beyond the critical distortion threshold (below which the Shannon lower bound is tight), except for the Gaussian source. 
It is worth noting that the Uniform distribution examined in \cite{rose94} is relatively more challenging compared to the aforementioned classical distributions. 
It exhibits a phase transition phenomenon, potentially leading to bifurcations depending on the chosen values of $D$. 
For this case, critical slope value $\lambda$ correspond to phase transitions. 
We consider the uniform source on interval $[-8,8]$ and conduct experiments with different discretization parameters, namely $K=20, 40, 80, 160$. The corresponding results are illustrated below in Figure 3.

\begin{figure}[ht]
    \centerline{\includegraphics[width=0.95\textwidth]{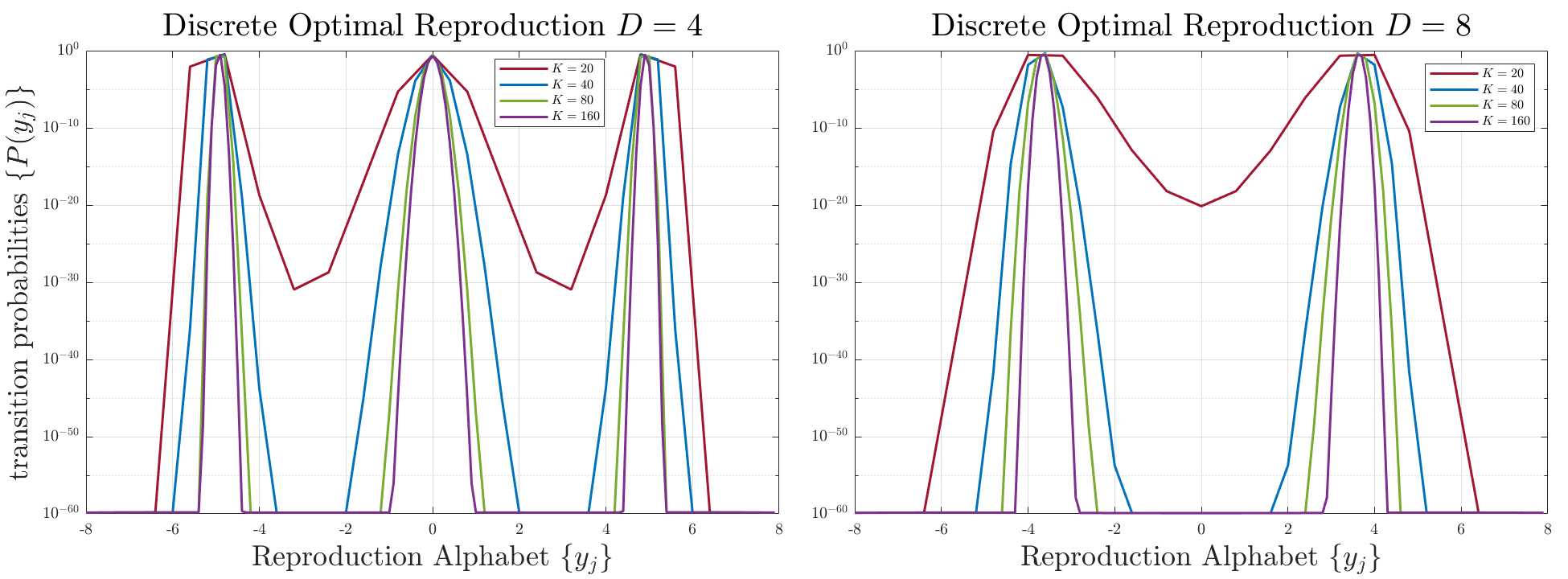}}
    \label{Fig: uniform}
   \caption{The discrete optimal reproduction  produced by the CBA algorithm for the cases of distortion $D=4$ (Left) and $D=8$ (Right).}
\end{figure}

From Figure 3, it can be observed that by increasing the discretization parameter $K$, the grid becomes finer, and the distribution of Y gradually approaches a discrete distribution. 
Meanwhile, our proposed CBA algorithm can still accurately compute the aforementioned situations. 
{As shown in \cite{rose94}, Rose} plots the graph of distortion versus the slope $\lambda$ on logarithmic scale to demonstrate the critical values where phase transitions occur.
In such cases, the traditional BA algorithm often proves to be inefficient due to the search of slope $\lambda$. 
However, our CBA algorithm can still swiftly provide solutions. 
In general terms, our algorithm exhibits an acceleration of approximately tens of times.
\begin{table}[ht] 
	\renewcommand\arraystretch{1.0}
	\centering
	\caption{Comparison between CBA and BA algorithms for the uniform source.} 
    \vspace{+.1in}
	\setlength{\tabcolsep}{1.5 mm}{
        \scalebox{1.0}{
		\begin{tabular}{c|c|c|c}
		\toprule
                & &\multicolumn{2}{c}{(Rate value, Number of iterations, Time in sec)} \\
			\hline 
                \multirow{10}{*}{CBA} & 
		      & $D=2$ & $D=4$ \\ 
                \cline{2-4}
                & $K=20$ & $(1.0602,\ 49756,\ 2.7308)$ & $(0.7366,\ 5683,\ 0.3085)$ \\
                & $K=40$ & $(1.0590,\ 103996,\ 6.9398)$ & $(0.7363,\ 23529,\ 1.56761)$ \\
                & $K=80$ & $(1.0585,\ 235964,\ 29.8931)$ & $(0.7361,\ 85857,\ 10.4445)$ \\
                & $K=160$ & $(1.0584,\ 581499,\ 91.5655)$ & $(0.7360,\ 252306,\ 63.2701)$ \\
                \cline{2-4}
                & & $D=8$ & $D=16$ \\ 
                \cline{2-4}
                & $K=20$ & $(0.4257,\ 2449,\ 0.1488)$ & $(0.1352,\ 9676,\ 0.6402)$ \\
                & $K=40$ & $(0.4244,\ 16187,\ 1.1517)$ & $(0.1352,\ 20338,\ 1.4591)$ \\
                & $K=80$ & $(0.4244,\ 24708,\ 2.5583)$ & $(0.1351,\ 61654,\ 6.5499)$ \\
                & $K=160$ & $(0.4243,\ 73139,\ 11.93991)$ & $(0.1351,\ 382922,\ 61.5829)$ \\
                \hline 
                \hline 
			\multirow{10}{*}{BA} 
                & & $D=2$ & $D=4$ \\ 
                \cline{2-4}
                & $K=20$ & $(1.0602,\ 59426,\ 2.7165e+03)$ & $(0.7366,\ 12793,\ 541.6412)$ \\
                & $K=40$ & $(1.0590,\ 57426,\ 3.3620e+03)$ & $(0.7363,\ 26806,\ 1.5604e+03)$ \\
                & $K=80$ & $(1.0585,\ 257193,\ 2.6684e+04)$ & $(0.7361,\ 100314,\ 1.3824e+04)$ \\
                & $K=160$ & $(1.0584,\ 941873,\ 1.1688e+05)$ & $(0.7360,\ 1227635,\ 1.7510e+05)$ \\
                \cline{2-4}
                & & $D=8$ & $D=16$ \\ 
                \cline{2-4}
                & $K=20$ & $(0.4257,\ 2899,\ 29.0791)$ & $(0.1352,\ 32626,\ 102.5707)$ \\
                & $K=40$ & $(0.4244,\ 55536,\ 241.6922)$ & $(0.1352,\ 54914,\ 457.8094)$ \\
                & $K=80$ & $(0.4244,\ 49010,\ 1.3831e+03)$ & $(0.1353^*,\ 18196,\ 1.7348e+03)$ \\
                & $K=160$ & $(0.4243,\ 111626,\ 1.2205e+03)$ & $(0.1351,\ 60398,\ 1.0436e+03)$ \\
		\bottomrule
	\end{tabular}}
 
\vspace{+.1in}

\footnotesize{{Notes: a) Columns 3-4 are the computed rate (in bits), the number of iterations and the average computing time (in seconds). b) The BA algorithm generally takes about $70$ trials to find the corresponding $\lambda$ for a given distortion $D$ with high accuracy. In the column for the BA algorithm, the time is measured including all the trials, whereas the number of iterations is only for the last trial with the found $\lambda$. c) The stop condition is that the decrease from the previous step is below $10^{-13}$. d) $*$ marks inaccurate results.}}}
\label{table_UNI} 
\end{table}

As depicted in Table \ref{table_UNI}, we present the computational time, iteration count, and computed rate values generated by both the CBA and BA algorithms.
It is obvious that the proposed CBA algorithm demonstrates a remarkable {speed-up ratio} of up to 100, with the iteration number typically being lower than that of the BA algorithm \footnote{This comparison pertains to the accurate computation of the slope in the BA algorithm.}.
Significantly, while the BA algorithm may occasionally yield results with fewer iterations, it is important to highlight that, in such instances, it falls short of delivering a {sufficiently} accurate solution unless an exceedingly high accuracy threshold is employed.
Therefore, in this experimental context, we have enhanced the accuracy threshold to $10^{-13}$ in order to ensure that both algorithms produce accurate results. 
This adjustment is particularly necessary for the uniform source, due to its phase transition phenomena. 
In fact, for this case, the slope undergoes significant variations across distortion values, and even slight perturbations in the slope value $\lambda$ can result in substantial errors when computing the RD function for a given distortion. 
By further enhancing the grid discretization, the CBA algorithm proves to be faster and more efficient than the BA algorithm in terms of computation time. 
Additionally, when comparing the computed rate values, the CBA algorithm also demonstrates superior accuracy in addressing the aforementioned issues compared to the BA algorithm.

Finally, it is noteworthy that as the grid discretization is intensified, the RD curve undergoes refinement \cite{rose94}, progressively approaching the characteristics of continuous cases. 
Consequently, for computation of the RD function in continuous scenarios, a higher degree of precision is often required, potentially giving rise to numerical challenges. 
In such circumstances, the utilization of the CBA algorithm may offer a more accurate, direct, and efficient approach.

\section{Conclusion} \label{Sec_5_conclusion}
In this paper, we propose a novel modification of the Blahut-Arimoto (BA) algorithm, named the Constrained-Blahut-Arimoto (CBA) algorithm, presenting a versatile framework for directly computing the Rate-Distortion (RD) and Distortion-Rate (DR) functions. 
Within this framework, we rigorously establish that the CBA algorithm exhibits a convergence rate of at least $O(1/n)$ and the computational complexity is estimated to be $O\left(\frac{MN\log N}{\varepsilon}(1+\log|\log \varepsilon|)\right)$ for achieving an $\varepsilon$-approximation to the optimal solution.
Numerical experiments convincingly demonstrate the effectiveness of the CBA algorithm, showcasing its superiority over the BA algorithm and other existing algorithms in the literature.
It can be envisioned that the CBA algorithm and the key ideas behind it may lead to a series of further applications in lossy compression as well as other problems in information theory and machine learning.

\section*{Acknowledgments}
This work was partially supported by National Key Research and Development Program of China (2018YFA0701603) and National Natural Science Foundation of China Grants (12271289 and 62231022). 

\bibliographystyle{abbrv}
\bibliography{RD_REF.bib}

\end{document}